\documentstyle[12pt]{article}
\newcommand{\be}{\begin{equation}}
\newcommand{\ee}{\end{equation}}
\newcommand{\bear}{\begin{eqnarray}}
\newcommand{\ear}{\end{eqnarray}}
\date{}
\renewcommand{\theequation}{\arabic{section}.\arabic{equation}}
\newcommand{\GeV}{\mbox{$\;$GeV}}

\newcommand{\M}{{\cal M}}

\newcommand{\PP}{I\!\!P}

\newcommand{\bm}[1]{\mbox{\bf #1}}  
\renewcommand{\vec}{\bm}
\newcommand{\grgl}{\:\hbox to -0.2pt{\lower2.5pt\hbox{$\sim$}\hss}
           {\raise3pt\hbox{$>$}}\:}
\newcommand{\klgl}{\:\hbox to -0.2pt{\lower2.5pt\hbox{$\sim$}\hss}
           {\raise3pt\hbox{$<$}}\:}

\begin{document}
\begin{flushright}
HD--THEP--96--06\\
DAMTP--96--51
\end{flushright}
\quad\\
\begin{center}
{\bf\LARGE SOME TESTS}\\
\bigskip
{\bf \LARGE FOR THE HELICITY STRUCTURE}\\
\bigskip
{\bf\LARGE OF THE POMERON IN $ep$ COLLISIONS}\\
\vspace{.5cm}
by\\
\vspace{.5cm}
T. Arens, O. Nachtmann\\
\bigskip
Institut  f\"ur Theoretische Physik\\
Universit\"at Heidelberg\\
Philosophenweg 16, D-69120 Heidelberg, FRG\\
\vspace{.4cm}
and\\
\vspace{.4cm}
M. Diehl, P. V. Landshoff\\
\bigskip
Department of Applied Mathematics and Theoretical Physics\\
University of Cambridge\\
Cambridge CB3 9EW, England
\vspace{.5cm}
\begin{abstract}
We discuss the reaction $ep\to e\tilde pX$ in the one-photon
exchange approximation, where it is in essence the reaction
$\gamma^*p\to\tilde pX$. A large rapidity gap is required
between the particle or
particles of the proton remnant $\tilde p$ and those
of $X$. We define a suitable azimuthal angle $\varphi$ between
a leptonic and a hadronic plane. The dependence of the cross
section on $\varphi$ is given explicitly and can be used to extract
cross sections and interference terms for the reaction
$\gamma^*p\to\tilde pX$ corresponding to the various helicities
of the virtual photon $\gamma^*$. The interference terms can be used
to test models for the large rapidity gap events in a sensitive
way. We discuss in detail models with factorizing Pomeron exchange
and in particular the Donnachie-Landshoff Pomeron model. We make
some remarks on soft colour exchange models and on possible effects
of QCD background vacuum fields. We conclude with a suggestion
to look for Odderon exchange in exclusive deep inelastic high
energy reactions like $\gamma^*p\to\tilde p\pi^0$ and
$\gamma^*p\to\tilde p\eta$.
\end{abstract}
\end{center}

\newpage

\section{Introduction} The idea to look for signals of ``hard''
diffractive reactions in hadron-hadron and lepton-hadron scattering
proved to be a very fruitful one \cite{1}-\cite{3}. This and the
discovery of ``rapidity gap events'' at HERA \cite{4},\cite{5} have
triggered intensive experimental and theoretical research in
``Pomeron'', i.e.\ diffractive physics. Topical questions which are
asked today are for instance the following: Is there a ``soft'' and a
``hard'' Pomeron? How are these concepts related? Is the ``Pomeron''
related to an ordinary Regge trajectory (cf.~e.g.~\cite{6}), to an
effective $C=+1$ ``photon'' exchange \cite{8}, to two-gluon exchange
\cite{7}, to a perturbative gluon ``ladder'' \cite{9} or to
nonperturbative QCD effects \cite{10}-\cite{15}?

In this article we will propose measurements of some angular
correlations in diffractive interactions initiated by a virtual photon
which may shed some light on the above questions. We will also propose
to look for signals of an ``Odderon'', the $C=-1$ partner of the
``Pomeron'', in these reactions. To be specific: We consider electron
or positron-proton scattering (Fig.~1): \be\label{1.1}
e^{\mp}(k)+p(p)\to e^{\mp}(k')+\tilde p(\tilde p) +X(p_X),\ee where
the 4-momenta are indicated in brackets. A large rapidity gap between
the group of hadrons $X$ and those of $\tilde p$ is required.  Further
details on our definition of a rapidity gap are given below.  The
``proton remnant'' $\tilde p$ can again be a proton or a diffractively
excited proton. The kinematics of deep inelastic $ep$ scattering is
well known (cf.~e.g. \cite{16}). In
 our paper  we want to draw attention to  the kinematic
similarity of reaction (\ref{1.1})
with one pion electroproduction:
\be\label{1.2}
e^-(k)+p(p)\to e^-(k')+p(\tilde p)+\pi^0(p_\pi).\ee
The detailed kinematic analysis of (\ref{1.2}) was done a long time
ago \cite{17}-\cite{19}. It can in essence be carried over to
(\ref{1.1}) and adapted to the HERA reference system. This we do
in sect.~2 of our paper. We consider, as usual, the reaction
(\ref{1.1}) in the one-photon exchange approximation, where we
have a quasi two-body process
\bear\label{1.3}
&&\gamma^*(q)+p(p)\to\tilde p(\tilde p)+X(p_X)\nonumber\\
&&(q=k-k').\ear
The contribution from $Z$-exchange could easily be incorporated.
However, for the range of $Q^2=-q^2$ below 100 GeV$^2$ where
rapidity gap events have been studied at HERA so far, $Z$-exchange
is unimportant, and thus we will neglect it in the following.

As we will see, there is an interesting azimuthal angle
$\varphi$ between
suitably defined lepton and hadron planes. The dependence of
the cross section for reaction (\ref{1.1}) on this angle can be
used to disentangle the cross sections and interference terms
for the reaction (\ref{1.3}) corresponding to the various
helicities of the virtual photon $\gamma^*$. This will be
explained in detail in sect.~3.  where the main formula is
(\ref{3.28}). This formula makes the dependence of the cross
section on $\varphi$ completely explicit, with terms proportional
to 1, $\cos\varphi$, $\sin\varphi$ and $\cos 2\varphi$, where the
coefficients depend only on the other kinematic variables.
In sect.~4 we will have a look
at reaction (\ref{1.3}) in the rest system of the group $X$
of final-state hadrons. We will show that nonzero values
for certain interference terms can only occur if the
``Pomeron'' carries nonzero helicity. We will then make some
remarks on models for rapidity gap events in connection with
such interference terms in sect.~5. In sect.~6 we will discuss
possible ``Odderon'' contributions in inelastic electron-proton
scattering. Sect.~7 contains our conclusions. In appendix A
we discuss some kinematical questions. In appendix B we treat in
detail some issues for the case where $\tilde p$ in (\ref{1.1}),
(\ref{1.3}) is a diffractively excited proton. In appendix C we show
how a measurement of the interference terms through the
$\varphi$-dependence of the cross section as discussed in sect.~3
can be used  to obtain bounds on the cross section for
longitudinally polarized photons in (\ref{1.3}). This
could be quite useful since a direct measurement of the
longitudinal cross section - or rather a disentangling of the
longitudinal and transverse cross sections - is notoriously difficult.

\section{Production of rapidity gap events as quasi two-body
$\gamma^*-p$ scattering reactions}
\setcounter{equation}{0}

We will assume that the internal structure of the final
state hadronic systems $X$ and $\tilde p$ in (\ref{1.1})
is not resolved and that only the 4-momenta $p_X$ and $\tilde p$
are considered as observables. We have, however, to place some
requirements on the criteria used for defining $X$ and $\tilde p$
as specified below. For a given c.m.\ energy $\sqrt s$ the final
state of (\ref{1.1}) is then determined completely by 7 variables,
for which we can e.g. take the 3-momenta $\vec k',\tilde {\vec p}$ in
a suitable system  and $\tilde m^2=\tilde p^2$, the invariant
mass squared of the proton remnant. We collect the definitions
of our kinematic variables and some useful formulae for them
in tables 1,2. We always neglect the electron mass but
keep the proton mass.

Let us now consider the reaction (\ref{1.1}) in the rest frame
of the original proton. We choose a convenient coordinate system
there: The 3-axis in the direction of the $\gamma^*$ momentum
$\vec q$, the 1,2 axes such that $\vec p_X$ is in the 1-3 plane
(Fig.~2). In this system we define the angle $\varphi$ as the
azimuthal angle of the 3-momentum vector of the incoming
electron or positron. We have then:
\bear\label{2.1}
&&\sin\varphi=\frac{\vec q\cdot[(\vec q\times \vec p_X)\times
(\vec q\times \vec k)]}{|\vec q||\vec q\times \vec p_X||
\vec q\times \vec k|}\Biggr|_R,\nonumber\\
&&\cos\varphi=\frac{(\vec q\times \vec p_X)\cdot
(\vec q\times \vec k)}{|\vec q\times \vec p_X||
\vec q\times \vec k|}\Biggr|_R, \nonumber\\
&&0\leq\varphi<2\pi.\ear
Here and in the following the index $R$ indicates quantities
referring to the proton rest system.

The definition of $\varphi$ given in (\ref{2.1}) is the
exact analogue of the definition of the corresponding
angle for reaction (\ref{1.2}) used in \cite{17}-\cite{19}.
For the case of one
hadron semi-inclusive electroproduction and jet production the
analogous angle is well known \cite{201}. It has been used by experimentalists 
in fixed target experiments (cf.~e.g.~\cite{200}) and in the context of HERA 
physics in a first analysis of H1 data in \cite{202}.

One of the main points we want to make is now the following. If
the systems $X$ and $\tilde p$ in (\ref{1.1})
are defined by criteria
involving only \underbar{hadronic}  variables, no leptonic
ones, then the dependence of the cross section for reaction
(\ref{1.1}) on $\varphi$  (\ref{2.1}) can be made
completely explicit
and can be used to measure cross sections and interference
terms for definite helicities of the virtual photon $\gamma^*$.
Examples for ``good'' criteria to define $X$ and $\tilde p$
are as follows:

Criterion I: We require $\tilde p$ to be a single proton and ask
it to carry most of the energy of the original proton in the HERA
system. To be precise, we require
\be\label{2.100}
\xi\leq\xi_0\ll 1,\ee
where $\xi$ is defined in table 1 and $\xi_0$
is some fixed number. Kinematics guarantees then a large
rapidity gap between $\tilde p$ and all the particles
of $X$.

Criterion II: We require $\tilde p$ to be a single proton. For $X$
we take the set of all final states with arbitrary number $n$
of particles of momenta $p_i$ $(i=1,...,n)$ such that
\bear\label{2.2}
&&\sum_{i=1}^np_i=p_X,\nonumber\\
&&p_X\cdot p_i\leq M^2_{cut}\ll p_X\cdot\tilde p\quad{\rm for\ all}
\quad i,\ear
where $M_{cut}$ is some fixed mass value. This means that in
the rest system of $X$ all energies of the particles composing
$X$ are limited to $M^2_{cut}/m_X$, which is a rotationally
invariant condition. Clearly, (\ref{2.2})
 is a  rapidity gap condition, since it  requires no
particles in the phase space between the particles of $X$
and $\tilde p$ (Fig.~3).

Criterion III: We allow the system $\tilde p$ to consist
of a single proton or of a group of particles, the debris of
a diffractively excited proton. We require the invariant mass $\tilde
m$ $(\tilde m^2=\tilde p^2)$ to be limited by some fixed value
$\tilde m_{cut}$:
\be\label{2.200}
\tilde m\leq\tilde m_{cut}.\ee
Furthermore we require as in I:
\be\label{2.201}
\xi\leq\xi_0\ll1.\ee

Criterion IV: For the system $X$ we choose the same requirements
as in Criterion II. But instead of requiring $\tilde p$ to be a single
proton we allow it to consist of a group of $\tilde n=1,2,...$
particles with momenta $\tilde p_j\ (j=1,...,\tilde n)$ such that
\bear\label{2.3}
&&\sum^{\tilde n}_{j=1}\tilde p_j=\tilde p,\nonumber\\
&&\tilde p\cdot \tilde p_j\leq \tilde M^2_{cut}\quad {\rm for\ all}
\ j.\ear
Here $\tilde M_{cut}$ is a fixed constant. Typically one would
choose $\tilde M_{cut}$ to be
 of order a few GeV in order to guarantee that one is in the
diffractive region.

The criteria I - IV above define $X$ and $\tilde p$
by use of invariants formed from \underbar{hadron}
4-momenta only.

With such criteria to define the systems of final states
$X$ and $\tilde p$, these can be treated like quasi-particles.
In particular, the leptonic vector $\vec k$ in Fig.~2 can rotate around
the axis $\vec q$ \underbar{without} affecting $X$ and $\tilde p$.
The hadronic and the leptonic variables, especially the azimuthal
angle $\varphi$, are in a sense decoupled. As we will see, this is
appropriate from a physics point of view, if we want to study
the dynamics of the $\gamma^*p$ reaction (\ref{1.3}). The criteria
I and III are, presumably, rather straightforward to implement
experimentally. The criteria II and IV require $X$ and $\tilde p$
to correspond to rotationally invariant sets of final states in
their respective rest systems, but we will not really use this
in the present paper. An important property guaranteed by all four
criteria is that in the rest frame of $X$ the set of final states is
invariant under rotations around the photon axis. This will be essential for our discussion in sect.~4.

In the remainder of this section we will relate the angle
$\varphi$ (\ref{2.1}) to quantities directly measurable in
the HERA system  where the 4-momenta of the original
proton and electron are given by
\be\label{2.4}
p_H=\left(\begin{array}{c}
E_{p,H}\\
\sqrt{E^2_{p,H}-m^2_p}\;\; {\bf\hat p}_H\end{array}\right),\quad
 k_H=\left(\begin{array}{c} E_{e,H}\\  -E_{e,H}\;\;{\bf\hat p}_H\end
{array}\right),\quad
{\bf\hat p}_H=\left(\begin{array}{c} 0\\ 0\\ 1\end{array}\right).\ee
Here and in the following quantities referring to the HERA
system carry an index $H$.

Let us define two 4-vectors $l,n$:
\bear\label{2.5}
l^\mu:&=&\frac{1}{m_p}\varepsilon^{\mu\nu\rho\sigma}
p_\nu q_\rho k_\sigma,
\nonumber\\
n^\mu:&=&\frac{1}{m_p}\varepsilon^{\mu\nu\rho\sigma}
p_\nu q_\rho p_{X\sigma}.
\ear
We have in the rest system of the original proton:
\bear\label{2.6}
l_R&=&\left(\begin{array}{c}
0\\ \vec q\times \vec k\end{array}\right)_R,\nonumber\\
n_R&=&\left(\begin{array}{c}
0\\ \vec q\times \vec p_X\end{array}\right)_R.\ear
From (\ref{2.1}) and (\ref{2.6}) we obtain easily
\bear\label{2.7}
\sin\varphi&=&\frac{\sqrt{\nu^2+Q^2} \, \varepsilon_
{\alpha\beta\gamma\delta} \,
k^\alpha p^\beta q^\gamma\tilde
p^\delta}{m_p\sqrt{-l^2}\sqrt{-n^2}},\nonumber\\
\cos\varphi&=&-\frac{l\cdot n}{\sqrt{-l^2}\sqrt{-n^2}}.\ear
These are covariant expressions for $\sin\varphi,\cos\varphi$.
To evaluate them in the HERA system we define an auxiliary vector
$\vec a_\bot$ by
\bear\label{2.8}
\vec a_\bot\Bigr|_H:&=&\Biggl\{y\left(1+\frac{2xm^2_p}{s-m^2_p}\right)
\tilde{\vec p}_\bot\nonumber\\
&&-\frac{1}{s-m_p^2}\left[-t-m^2_p+\tilde m^2+2xym^2_p+m_p^2\frac{4p_X
\cdot k}{s-m^2_p}\right] \, \vec q_\bot\Biggr\}_H.\ear
In the HERA system we have for the transverse momenta
\bear\label{2.9}
(\vec q_\bot+\vec k_\bot')_H&=&0,\nonumber\\
(\vec q_\bot-\tilde{\vec p}_\bot-\vec p_{X\bot})_H&=&0,\ear
\bear\label{2.10}
\vec q^2_\bot\Bigr|_H&=&Q^2\left[1-y-\frac{m^2_pQ^2}
{(s-m^2_p)^2}\right]\nonumber\\
&=&xy(s-m^2_p)\left[1-y-\frac{xy m^2_p}
{s-m^2_p}\right].\ear
This leads us to
\bear\label{2.11}
l_H&=&-\frac{s-m^2_p}{2m_p}\left(\begin{array}{c}
0\\ \vec q_\bot\times{\bf\hat p}\end{array}\right)_H,\nonumber\\
n_H&=&-\left(\begin{array}{l}
\frac{1}{m_p}(\vec q_\bot \times\tilde{\vec p}_\bot)\cdot \vec p\\
\frac {s-m_p^2}{2m_p} \, \vec a_\bot\times {\bf\hat p}+\frac{p^0}{m_p}
\, \vec q_\bot\times\tilde{\vec p}_\bot\end{array}\right)_H,\ear
\bear\label{2.12}
-l^2&=&\left(\frac{s-m_p^2}{2m_p}\right)^2\vec q^2_\bot\Bigr|_H,
\nonumber\\
-n^2&=&\left\{\left(\frac{s-m_p^2}{2m_p}\right)^2\vec a_\bot^2+
(\vec q_\bot\times\tilde{\vec p}_\bot)^2\right\}_H\nonumber\\
&=&\left(\frac{s-m^2_p}{2m_p}\right)^2\Bigl\{
y^2 \left[-t(1-x)+m^2_px(1-x)-\tilde m^2x \right]\nonumber\\
&&+y(s-m_p^2)^{-1} \left[(t+m^2_p-\tilde m^2)(m_X^2-t)+x((t+m_p^2-\tilde
m^2)^2-2m^2_pt \right. \nonumber\\
&&\left. -2m^2_pm_X^2) \right]-(s-m^2_p)^{-2}m^2_p(m_X^2-t)^2\Bigr\},\ear
\bear\label{2.13}
\sin\varphi&=&y^{1/2}\left(y+\frac{4xm_p^2}{s-m_p^2}
\right)^{1/2}(\vec q_\bot\times\tilde{\vec p}_\bot)\cdot{\bf\hat p}
\nonumber\\
&&\cdot|\vec q_\bot|^{-1}\left[\vec a^2_\bot+\frac{4m_p^2(
\vec q_\bot\times \tilde{\vec p}_\bot)^2}{(s-m^2_p)^2}\right]^{-1/2}
\Biggr|_H,\nonumber\\
\cos\varphi&=&(\vec q_\bot\cdot\vec a_\bot)\, |\vec q_\bot|^{-1}\left[
\vec a_\bot^2+\frac{4m_p^2(\vec q_\bot\times\tilde{\vec p}_\bot)^2}
{(s-m_p^2)^2}\right]^{-1/2}\Biggr|_H.\ear
These are the desired expressions for $\sin\varphi,\cos\varphi$
in terms of quantities directly measurable in the HERA system. At
HERA we have $s\simeq9\cdot 10^4\ {\GeV}^2$ and thus in the kinematic region
where rapidity gap events are studied today \cite{4},\cite{5}:
\be\label{2.14}
s-m^2_p\gg m_p^2,\tilde m^2,m_X^2,Q^2,(-t).\ee
In this limit, i.e.\ for $s\to\infty$, we get from (\ref{2.8}),
(\ref{2.13}):
\bear
&&\vec a_\bot|_H\to y\tilde{\vec p}_\bot|_H,\label{2.171}\\
&&\sin\varphi\to \sin\varphi'=
\frac{(\vec q_\bot\times\tilde{\vec p}_\bot)\cdot{\bf\hat p}}
{|\vec q_\bot||\tilde{\vec p}_\bot|}\Bigr|_H,\nonumber\\
&&\cos\varphi\to\cos\varphi'=
\frac{\vec q_\bot\cdot\tilde{\vec p}_\bot}{|\vec q_\bot
||\tilde{\vec p}_\bot|}\Bigr|_H.\label{2.15}\ear
The angle $\varphi'$ is  the
azimuthal angle between $\vec q_\bot$ and $\tilde{\vec p}_\bot$ in
the HERA system.
This is illustrated in Fig.~4. Let $\tilde\psi$ and $\psi'$
be the azimuthal angles of $\tilde{\vec p}_\bot$ and $\vec k_\bot'$ with
respect to some fixed axes 1,2 in the HERA system, where the
3-axis is defined by the momentum vector of the incoming proton.
We have then for large $s$ (cf.~(\ref{2.14})):
\be\label{2.150}
\varphi\approx\varphi'=\tilde\psi-\psi'+\pi.\ee
To assess how good this approximation is we observe that at finite
$s$ the replacement (\ref{2.171}) becomes bad for very small
$|\tilde{\vec p}_\bot|_H$. Indeed from (\ref{2.8}) we see that this
happens for
\be\label{2.191}
|\tilde{\bm p}_\bot|_H\klgl\frac{|\vec q_\bot|_H}{y(s-m^2_p)}
\cdot(-t-m^2_p+\tilde m^2).\ee
Now we have $|\tilde{\bm p}_\bot|^2_H\approx |t|$ with
corrections of order $x|t|$ and $x^2 m^2_p$, and
$|\vec q_\bot|^2_H=Q^2(1-y) \cdot (1+O(xm^2_p/s))$ (cf.~(\ref{2.10})). From
this we find that the relative error which is made by using
 $\sin\varphi'$
 instead of $\sin\varphi$ is of order
\be\label{2.192}
O\left( \frac{|\tilde{\bm p}_\bot |_H}{Q} \cdot \frac{m^2_p,\tilde m^2,
  |t|}{ys} \right) = 
O\left( x\cdot \frac{m^2_p,\tilde m^2, |t|}{Q\, |\tilde{\bm p}_\bot |_H}
\right)
 .\ee
Thus small $x$ helps to keep the corrections small. On the other hand,
very small $|\tilde{\vec p}_\bot|_H$, which corresponds to very small
$|t|$, can lead to sizable corrections, as already noted above. We
certainly recommend that experimentalists use the exact definitions of
 $\sin\varphi $, $\cos\varphi$ (\ref{2.13}) whenever possible.

Coming back to the general case, we note that the angle $\varphi$
can only be defined unambiguously if the conditions
\be\label{2.16}
-l^2\not=0,\ee
\be\label{2.17}
-n^2\not=0\ee
are satisfied (cf.~(\ref{2.7})). We see from (\ref{2.12}) that
(\ref{2.16}) requires
\be\label{2.18}
\vec q_\bot|_H=-{\vec k}_\bot'|_H\not=0,\ee
i.e.\ the scattered electron must have nonzero transverse
momentum. Condition (\ref{2.17}) requires for $s\to\infty$ that
\be\label{2.19}
\tilde{\vec p}_\bot|_H\not=0,\ee
i.e.\ that the proton remnant has nonzero transverse momentum.
The exact meaning of $n^2\not=0$ for the case of arbitrary finite
$s$ is discussed in appendix A.

\section{Cross sections and interference terms for definite
$\gamma^*$ helicities in the proton rest system}
\setcounter{equation}{0}
Let us consider now in more detail the electroproduction
reaction
\be\label{3.1}
e^\mp(k)+p(p)\to e^\mp(k')+p(\tilde p)
+a_1(p_1)+...+a_n(p_n).\ee
Here we assume that the original proton is unpolarized, whereas
we allow arbitrary polarization for the incoming electron or positron.
The covariant density matrix of the initial $e^-$ or $e^+$
is then given by
\be\label{3.100}
(k\!\!\!/ \pm m_e)\, \frac{1+\gamma_5 s\!\!\!/ }{2},\ee
where  $s^\mu$ is
the $e^\mp$ polarization vector which satisfies:
\bear
s^\mu k_\mu&=&0,\label{3.101}\\
-1\leq s^\mu s_\mu&\leq& 0.\label{3.102}\ear
For $m_e\to0$ we can parametrize $s$ as follows:
\bear\label{3.2}
&&s\to r_L\frac{k}{m_e}+\left(\begin{array}{c} 0\\ \vec r_\bot
\end{array}\right),\nonumber\\
&&\vec r_\bot\vec k=0,\nonumber\\
&&r^2_L+\vec r_\bot^2\leq 1.\ear
Here the longitudinal and transverse polarizations are described by
$r_L$ and $\vec r_\bot$, respectively, with $r_L=+1(-1)$ corresponding
to a right(left)-handed electron or positron. We assume  in this
section that the proton remnant is again a single proton. The
more general case of the remnant being a group of particles is treated
in appendix B. The hadrons $a_1,...,a_n$ $(n=1,2,..)$ form
the quasiparticle $X$, where
\be\label{3.3}
\sum_{i=1}^n p_i=p_X.\ee

As is well known, the cross section for reaction (\ref{3.1}) can be
written as a product of a lepton and a hadron tensor
\be\label{3.4}
d\sigma(e^\mp p\to e^\mp pX)=\frac{4m_p}{s-m^2_p}\left(\frac{\alpha}
{Q^2}\right)^2l^{\nu\mu}W^{(X)}_{\mu\nu}\frac{d^3k'}{{k'}^0}\frac
{d^3\tilde p}{\tilde p^0},\ee
where (with the convention $\varepsilon_{0123}=1$):
\bear\label{3.5}
&&l^{\nu\mu}={k'}^\nu k^\mu+k^\nu {k'}^\mu-g^{\nu\mu}
(kk'-m^2_e)+im_e\, \varepsilon^{\nu\mu\alpha\beta}q_\alpha s_\beta\nonumber\\
&&\stackrel{\scriptstyle m_e\to 0}{\longrightarrow}
{k'}^\nu k^\mu +k^\nu{k'}^\mu
-g^{\nu\mu}(kk')+ir_L \, \varepsilon^{\nu\mu\alpha\beta}q_\alpha k_\beta,\ear
\bear\label{3.6}
\!\!\!\!\!\!&&W_{\mu\nu}^{(X)}(p,q,\tilde
p)=\frac{1}{4m_p}\sum_n\int_{p_1}...\int_{p_n}\prod^n_{i=1}
\frac{d^3p_i}{(2\pi)^32p_i^0}\nonumber\\
\!\!\!\!\!\!&&\chi_n(p_1,...,p_n,p,q,\tilde p)\, \delta(p+q-\tilde p-
\sum^n_{i=1}p_i)\nonumber\\
\!\!\!\!\!\!&&\sum_{spins}\,\!\! ^\prime <p(p)|J_\mu(0)|a_i(p_i),
p(\tilde p)\ {\rm out}>
<a_i(p_i),p(\tilde p)\ {\rm out}|J_\nu(0)|p(p)>.\ear
Here and in the following $\sum_{spins}'$ indicates the average and sum
over the spin states  of the initial and final state particles,
respectively.
The hadronic electromagnetic current is denoted by $eJ_\mu$ and
the functions $\chi_n$ are inserted as characteristic
functions being 0 outside and 1 inside
the region in phase space defining the set of
final states $X$. The conventions for the normalization of states,
for $\gamma$-matrices etc., are as in \cite{16}, in particular
we set for a spinless particle $a(p)$:
\be\label{3.600}
<a(p')|a(p)>=(2\pi)^3 2p^0\delta^3(\vec p'-\vec p).\ee

The next step is to consider (\ref{3.1}) as a reaction
where a virtual photon $\gamma^*$ with density matrix
$l^{\nu\mu}$ is produced which interacts then with the
proton. We analyse now the helicity content of the density
matrix $l^{\nu\mu}$ in the proton rest system. For this we
define a vierbein of polarization vectors $\varepsilon_\alpha
\ (\alpha=0,1,2,3)$ as follows:
\bear\label{3.7}
\varepsilon_0&=&\frac{(p\cdot q)q+Q^2p}{\sqrt{Q^2
[(p\cdot q)^2+m^2_pQ^2]}},
\nonumber\\
\varepsilon_1&=&\frac{\eta}{\sqrt{-\eta^2}},\nonumber\\
\varepsilon_2&=&\frac{n}{\sqrt{-n^2}},\nonumber\\
\varepsilon_3&=&\frac{q}{\sqrt{-q^2}},\ear
where $n$ is defined in (\ref{2.5}) and
\be\label{3.8}
\eta=p_X-\frac{(p\cdot p_X)}{m_p^2}p+\frac{m_p^2
(q\cdot p_X)-(p\cdot q)(p\cdot p_X)}{(p\cdot q)^2
+m_p^2Q^2}\left[q-\frac{(p\cdot q)}{m^2_p}p\right].\ee
In the rest system of the original proton with the
coordinate axes as defined in Fig.~2 we have
\bear\label{3.9}
\varepsilon_0\Bigr|_R&=&\frac{1}{Q}\left(\begin{array}
{c} \sqrt{\nu^2+Q^2}\\ 0\\ 0\\ \nu\end{array}\right),
\nonumber\\
\varepsilon_1\Bigr|_R&=&\left(\begin{array}
{c} 0\\ 1\\ 0\\ 0\end{array}\right),
\nonumber\\
\varepsilon_2\Bigr|_R&=&\left(\begin{array}
{c} 0\\ 0\\ 1\\ 0\end{array}\right),
\nonumber\\
\varepsilon_3\Bigr|_R&=&\frac{1}{Q}\left(\begin{array}
{c} \nu\\ 0\\ 0\\ \sqrt{\nu^2+Q^2}\end{array}\right).
\ear
From this we find easily
\bear\label{3.10}
&&(\varepsilon_\alpha\cdot \varepsilon_\beta)=g_{\alpha\beta}
,\nonumber\\
&&\det(\varepsilon_\alpha,\varepsilon_\beta,\varepsilon
_\gamma,\varepsilon_\delta)
=\varepsilon_{\alpha\beta\gamma\delta}.\ear
The virtual photon $\gamma^*$ can have polarization vectors
$\varepsilon_0$ and $\varepsilon_{1,2}$ corresponding to longitudinal
and transverse polarizations, respectively, in the system $R$.
The polarization vectors for right and left circular polarization
are defined as
\be\label{3.11}
\varepsilon_\pm=\mp\frac{1}{\sqrt2}(\varepsilon_1\pm i
\varepsilon_2).\ee

It is now straightforward to express the density matrix
$l^{\nu\mu}$ in terms of the vectors $\varepsilon_\alpha$. We
get
\bear\label{3.12}
l^{\nu\mu}&=&\frac{Q^2}{1-\varepsilon}\Biggl\{\frac{1}{2}
\left(\varepsilon_+^\nu
\varepsilon^{\mu*}_++\varepsilon^\nu_-\varepsilon^{\mu*}_-\right)
+\varepsilon\, \varepsilon_0^\nu\varepsilon^{\mu*}_0\nonumber\\
&&-\varepsilon\cos 2\varphi\, \frac{1}{2} \left(\varepsilon^\nu_-
\varepsilon_+^{\mu*}
+\varepsilon^\nu_+\varepsilon^{\mu*}_-\right)+\varepsilon\sin
2\varphi\, \frac{1}{2i} \left(\varepsilon^\nu_-\varepsilon_+^{\mu*}
-\varepsilon^\nu_+\varepsilon^{\mu*}_-\right)\nonumber\\
&&-\sqrt{\varepsilon(1+\varepsilon)}\cos\varphi\, \frac{1}{2}
\left[\varepsilon^\nu_0(\varepsilon_+-\varepsilon_-)^{\mu*}+
(\varepsilon_+-\varepsilon_-)^\nu\varepsilon_0^{\mu*}\right]\nonumber\\
&&+\sqrt{\varepsilon(1+\varepsilon)}\sin\varphi\, \frac{1}{2i} 
\left[\varepsilon^\nu_0(\varepsilon_++\varepsilon_-)^{\mu*}-
(\varepsilon_++\varepsilon_-)^\nu\varepsilon_0^{\mu*}\right]\nonumber\\
&&+r_L\sqrt{1-\varepsilon^2}\, \frac{1}{2} \left[\varepsilon_+^\nu\varepsilon_+
^{\mu*}-\varepsilon_-^\nu\varepsilon_-^{\mu*}\right]\nonumber\\
&&-r_L\sqrt{\varepsilon(1-\varepsilon)}\cos\varphi\, \frac{1}{2} 
\left[\varepsilon^\nu_0(\varepsilon_++\varepsilon_-)^{\mu*}+
(\varepsilon_++\varepsilon_-)^\nu\varepsilon_0^{\mu*}\right]\nonumber\\
&&+r_L\sqrt{\varepsilon(1-\varepsilon)}\sin\varphi\, \frac{1}{2i}
\left[\varepsilon^\nu_0(\varepsilon_+-\varepsilon_-)^{\mu*}-
(\varepsilon_+-\varepsilon_-)^\nu\varepsilon_0^{\mu*}\right]\Biggr\}.
\ear
Here the quantity $\varepsilon$ is the usual ratio of longitudinal
to transverse polarization strength of the $\gamma^*$ (cf.
table 1).

We define now the cross sections and interference terms for the
absorption of virtual photons $\gamma^*$ of definite helicity
in the system $R$:
\be\label{3.13}
\sigma_{mn}^{(X)}:=\frac{4\pi^2\alpha}{K}\varepsilon^
{*\mu}_mW_{\mu\nu}^{(X)}\varepsilon^\nu_n \qquad (m,n=0,\pm1),\ee
where we use Hand's convention \cite{20} for the flux factor with
$K$ as defined in table 1. It is clear that the matrix of these
cross sections $(\sigma_{mn}^{(X)})$ is hermitian and positive
semi-definite. This means that we must have:
\be\label{3.14}
\sigma_{mn}^{(X)}=\sigma_{nm}^{(X)*},\ee
and for an arbitrary complex vector $c=(c_m)$:
\be\label{3.15}
\sum_{m,n} c^*_m\sigma_{mn}^{(X)} c_n\geq 0.\ee
These properties of the cross section matrix $(\sigma_{mn}^{(X)})$
are discussed further in Appendix C. There we also show how the
positivity condition (\ref{3.15}) may be used to get information
on the longitudinal cross section $\sigma_{00}^{(X)}$ ---
a quantity which is notoriously difficult to measure experimentally
--- from the measurements of the interference terms
$\sigma_{mn}^{(X)}(m\not=n)$.

Inserting (\ref{3.12}) and (\ref{3.13}) in (\ref{3.4})
we find
\bear\label{3.16}
d\sigma(e^\mp p\to e^\mp pX)&=&
\Gamma\Biggl\{\frac{1}{2}(\sigma_{++}^{(X)}+\sigma_{--}^{(X)})+\varepsilon
\sigma_{00}^{(X)}\nonumber\\
&&-\varepsilon\cos(2\varphi)\ {\rm Re}\sigma_{+-}^{(X)}\nonumber\\
&&+\varepsilon\sin(2\varphi)\ {\rm Im}\sigma_{+-}^{(X)}\nonumber\\
&&-\sqrt{\varepsilon(1+\varepsilon)}\cos\varphi\
{\rm Re}(\sigma_{+0}^{(X)}-\sigma_{-0}^{(X)})\nonumber\\
&&+\sqrt{\varepsilon(1+\varepsilon)}\sin\varphi\
 {\rm Im}(\sigma_{+0}^{(X)}
+\sigma_{-0}^{(X)})\nonumber\\
&&+r_L\sqrt{1-\varepsilon^2}\frac{1}{2}
(\sigma_{++}^{(X)}-\sigma_{--}^{(X)})
\nonumber\\
&&-r_L\sqrt{\varepsilon(1-\varepsilon)}\cos\varphi\ {\rm Re}
(\sigma_{+0}^{(X)}+\sigma_{-0}^{(X)})
\nonumber\\
&&+r_L\sqrt{\varepsilon(1-\varepsilon)}\sin\varphi\ {\rm Im}
(\sigma_{+0}^{(X)}-\sigma_{-0}^{(X)})\Bigr\}\nonumber\\
&&\cdot\frac{m_p}{(pk')}\frac{d^3k'}{{k'}^0}\frac
{d^3\tilde p}{\tilde p^0},\ear
where $\Gamma$ is the conventional flux factor of
virtual photons (cf.~\cite{19}):
\be\label{3.17}
\Gamma=\frac{\alpha}{2\pi^2}\frac{(p\cdot k')}{(p\cdot k)}\frac{W^2-m_p^2}
{2m_pQ^2}\frac{1}{1-\varepsilon}.\ee
So far the analysis is completely general, valid for any
set of final states $X$.

Now we will consider the case that the final state $X$ is defined
according to criterion I or II of sect.~2.
Choosing criterion I (\ref{2.100})
we have to set for the functions $\chi_n$ in (\ref{3.6})
\be\label{3.171}
\chi_n^{(I)}(p_1,...,p_n,p,q,\tilde p)=
\Theta\left(\xi_0-\frac{(p-\tilde p)\cdot q}{p\cdot q}\right).\ee
Here $\Theta(\cdot)$ is the usual step function. For criterion II
the relevant condition is given in (\ref{2.2}). This is implemented
by setting for the functions $\chi_n$ in (\ref{3.6}):
\be\label{3.18}
\chi_n^{(II)}(p_1,...,p_n,p,q,\tilde p)=
\prod^n_{j=1}\Theta\left[M^2_{cut}-p_j\cdot (p+q-\tilde p)\right].\ee
In both cases, (\ref{3.171}) and (\ref{3.18}) we can
treat $X$ as a quasi particle with $W_{\mu\nu}^{(X)}$ depending
only on the 4-momenta $p,q,\tilde p$:
\be\label{3.19}
W^{(X)}_{\mu\nu}=W^{(X)}_{\mu\nu}(p,q,\tilde p).\ee
We note that the interference terms
$\sigma_{mn}^{(X)}$  $(m\not=n)$ must vanish for $t=-|t|_{min}$ and
$t=-|t|_{max}$ (cf.~Appendix A). For these $t$ values the reaction
$\gamma^* p\to pX$ is collinear in the  original proton
rest system $R$. With any of the criteria I or II the final  state
$pX$ is then rotationally invariant around the collision axis, which
implies by an elementary angular momentum conservation argument:
 \be\label{3.170}
\sigma_{mn}^{(X)}=0\quad{\rm for}\quad t=-|t|_{min},\quad
{\rm and}\quad t=-|t|_{max}\quad{\rm if}\quad m\not=n.\ee
Due to (\ref{3.170}) the $\varphi$-dependent terms in (\ref{3.16}) vanish
for the collinear situation and thus the fact that
 the angle $\varphi$ becomes
undefined there (cf.~(\ref{2.17})) causes no problems.

The functions $\chi_n^{(I,II)}$, (\ref{3.171})
and (\ref{3.18}), are parity-invariant. Then
parity invariance of the strong interactions gives the
condition:
\be\label{3.20}
W_{\mu\nu}^{(X)}(p,q,\tilde p)=W^{(X)\mu\nu}
(p',q', \tilde p'),\ee
where the parity transformed 4-momenta are:
\be\label{3.21}
{p'}^\mu=p_\mu,\ {q'}^\mu=q_\mu,\ \tilde{p}' {}^\mu=\tilde p_\mu.\ee
Furthermore we have from (\ref{3.7})-(\ref{3.11})
\be\label{3.22}
\varepsilon_m^\mu(p',q',\tilde p')=(-1)^m\varepsilon_{-m\mu}
(p,q,\tilde p) \qquad (m=0,\pm1).\ee
From (\ref{3.13}) and (\ref{3.19}) we find that the cross
sections $\sigma_{mn}^{(X)}$ are Lorentz-invariant functions
of the momenta $p,q,\tilde p$ and thus only functions of the
invariants we can form from $p,q,\tilde p$. There are 4 such
invariants, all parity even, a convenient set being
(cf.~table 1)
\be\label{3.23}
x,Q^2,\xi,t.\ee
We get then
\be\label{3.24}
\sigma_{mn}^{(X)}=\sigma^{(X)}_{mn}(x,Q^2,\xi,t).\ee
From parity invariance (\ref{3.20}) and (\ref{3.22}) we
get now:
\be\label{3.25}
\sigma_{mn}^{(X)}(x,Q^2,\xi,t)=(-1)^{m+n}\sigma_{-m,-n}^{(X)}
(x,Q^2,\xi,t).\ee
From (\ref{3.14}) and (\ref{3.25}) we find the following
relations:
\bear\label{3.26}
\sigma_{++}^{(X)}&=&\sigma_{--}^{(X)},\nonumber\\
\sigma_{+-}^{(X)}&=&\sigma_{-+}^{(X)} \:\, = \:\,
\sigma_{+-}^{(X)*},\nonumber\\ 
\sigma_{+0}^{(X)}&=&-\sigma_{-0}^{(X)}.\ear

The last step is to insert (\ref{3.26}) into the
general formula (\ref{3.16}) and to express the phase space
element in terms of the variables
\[x,Q^2,\xi,t,\varphi,\tilde\psi\]
where $\tilde\psi$ is the azimuthal angle of $\tilde {\vec p}$ in the HERA
system (Fig.~4) with respect to a fixed laboratory frame. We get:
\be\label{3.27}
\frac{d^3k'}{{k'}^0}\frac{d^3\tilde p}{\tilde p^0}=\frac{y}{4x}\left(1+
\frac{4x^2m_p^2}{Q^2}\right)^{-1/2}dx\, dQ^2\, d\xi\, 
 dt\, d\varphi\,  d\tilde\psi,\ee
\bear\label{3.28}
&&\frac{d^6\sigma(ep\to epX)}{dx\, dQ^2\, d\xi\,  dt\, d\varphi\,  d\tilde\psi}
=\frac{1}{2\pi} 
\tilde\Gamma\Biggl\{\sigma_{++}^{(X)}+\varepsilon\sigma_{00}^{(X)}
-\varepsilon\cos(2\varphi)\sigma_{+-}^{(X)}\nonumber\\
&&-2\sqrt{\varepsilon(1+\varepsilon)}\cos\varphi\
{\rm Re}\sigma_{+0}^{(X)}
+2r_L\sqrt{\varepsilon(1-\varepsilon)}\sin\varphi\ {\rm Im}\sigma_{+0}
^{(X)}\Biggr\}.\ear
The flux factor $\tilde\Gamma$ is given by
\bear\label{3.29}
&&\tilde\Gamma=\frac{\pi m_py^{3/2}}{2x(pk')}\left(y+
\frac{4xm_p^2}{s-m^2_p}\right)^{-1/2}\cdot\Gamma\nonumber\\
&&=\frac{\alpha}{2\pi(s-m_p^2)}\frac{m_pKy^{3/2}}{xQ^2(1-\varepsilon)}
\left(y+\frac{4xm^2_p}{s-m_p^2}\right)^{-1/2}.\ear
The right-hand side of (\ref{3.28}) is, of course,
independent of the overall azimuthal angle $\tilde\psi$ in the HERA
system. To measure the cross section (\ref{3.28}) and in particular
the $\varphi$ dependence we think of an experimental setup
where the scattered proton $p(\tilde p)$ is detected with
the help of a small angle forward detector having some azimuthal
coverage $\Delta\tilde\psi$. The relative
angle $\varphi'$ (cf.~Fig.~4) which is simply related to $\varphi$
(cf.~(\ref{2.13})-(\ref{2.150})) should then easily be measurable
since the ZEUS and H1 detectors have full azimuthal coverage for
the scattered lepton. Integrating (\ref{3.28}) over
$\tilde\psi$ in the interval $\Delta\tilde\psi$ we get
\bear\label{3.290}
\frac{d^5\sigma(ep\to epX)}{dx\, dQ^2\, d
\xi\,  dt\,  d\varphi}&=&\frac{\Delta\tilde\psi}{2\pi}
\tilde\Gamma
\Bigl\{\sigma_{++}^{(X)}+\varepsilon\sigma_{00}^{(X)}
-\varepsilon\cos(2\varphi)\sigma_{+-}^{(X)}\nonumber\\
&&-2\sqrt{\varepsilon(1+\varepsilon)}\cos\varphi\ {\rm Re}\
\sigma_{+0}^{(X)}\nonumber\\
&&+2r_L\sqrt{\varepsilon(1-\varepsilon)}\sin\varphi\ {\rm Im}\
\sigma_{+0}^{(X)}\Bigr\}.\ear

The case where the scattered proton is not detected and the
scattered lepton is accepted over the full azimuth corresponds
to integrating over $\varphi$ and $\tilde\psi$ in (\ref{3.28}). We
get then
\be\label{3.30}
\frac{d^4\sigma(ep\to epX)}{dx\, dQ^2\, d\xi\,
dt}=2\pi\tilde\Gamma(\sigma_{++}^{(X)} 
+\varepsilon\sigma^{(X)}_{00})\ee
which allows us to make easy contact with the conventional
notation for this 4-dimensional cross section \cite{4}, \cite{5}.
Let us define
\bear\label{3.31}
&&F_2^{D(4)}(x,Q^2,\xi,t):=\frac{Q^2(1-x)}{4\pi\alpha}
\left(1+\frac{4x^2m_p^2}{Q^2}\right)^{-3/2}(\sigma_{++}^{(X)}
+\sigma_{00}^{(X)}),\nonumber\\
&&R^{D(4)}(x,Q^2,\xi,t):=\frac{\sigma_{00}^{(X)}}{\sigma_{++}^{(X)}}.
\ear
We get then from (\ref{3.30}) the usual expression:
\bear\label{3.32}
\frac{d^4\sigma(ep\to epX)}{dx\, d
Q^2\, d\xi\,  dt}&=&\frac{4\pi\alpha^2}{xQ^4}
\Biggl\{1-y-\frac{m_p^2xy}{s-m^2_p}\nonumber\\
&&+\frac{1}{2}\left[y^2+\frac{4xym^2_p}{s-m^2_p}\right]
\left[1+R^{D(4)}\right]^{-1}\Biggr\}F_2^{D(4)}.\ear

The formula (\ref{3.28}) is the main result of this section. Note
that in (\ref{3.28}) the dependence of the cross section on
the azimuthal angle $\varphi$ is completely explicit. The
coefficients $\sigma_{++}^{(X)},\sigma_{00}^{(X)}$ etc. depend
only on $x,Q^2, \xi, t$ (cf.~(\ref{3.24})). The neat separation
of $\varphi$ from these latter kinematic variables which describe
the reaction $\gamma^*p\to pX$ is due to the fact that
(a) we chose to analyse the reaction (\ref{3.1}) in the proton
rest system $R$ with respect to $\vec q$ as 3-axis, and (b)
we imposed the criterion I (\ref{2.100}) or criterion II (\ref{2.2})
for defining $X$ which allowed
us to treat $X$ as a quasi particle. In this respect our analysis
differs in a fundamental way from the analysis presented in
\cite{21} where instead of the $\vec q$ axis in $R$ the axis
defined by the electron and proton beams in the HERA system
is used to define $X$ and the kinematic variables, in particular
an azimuthal angle. Not
surprisingly, in the latter case the kinematic dependence
on their azimuthal angle is coupled with the dynamic dependence
on the momentum transfer squared $t$ of reaction (\ref{1.3}). We
think it is fair to say that the analysis of \cite{21} may be more
easily implemented experimentally whereas the analysis proposed in
our paper requires more experimental work but - as we hope to show
below - has the promise of rendering more directly interesting
information on the dynamics of the diffractive reaction (\ref{1.3}),
namely on the helicity structure of the ``Pomeron''.

\section{The helicity structure of the reaction
$\gamma^*p\to pX$ in the rest system of $X$}
\setcounter{equation}{0}
In this section we propose to look at reaction (\ref{3.1})
in the rest system of $X$, denoted by $RX$ in the following.
We will define cross sections and interference terms for the
various helicities of $\gamma^*$ in the system $RX$.

We start by defining a coordinate system in $RX$ with the
$\gamma^*$ 3-momentum $\vec q|_{RX}$ as 3-axis
and the plane spanned by $\vec q|_{RX}$ and $\vec p|_{RX}$
as 1-3 plane (Fig.~5). We define a vierbein of polarization
vectors $\tilde\varepsilon_\alpha$ $(\alpha=0,1,2,3)$ related
to the system $RX$ as in (\ref{3.7}) but with the replacements:
\be\label{4.1}
p\to p_X,\ m_p\to m_X,\ p_X\to p.\ee
This leads to:
\bear\label{4.2}
\tilde\varepsilon_0&=&\frac{(p_X\cdot q)q+Q^2p_X}
{\sqrt{Q^2[(p_X\cdot q)^2+Q^2m^2_X]}},\nonumber\\
\tilde\varepsilon_1&=&\frac{\tilde \eta}{\sqrt{-\tilde\eta^2}},
\nonumber\\
\tilde\varepsilon_2&=&\frac{\tilde n}{\sqrt{-\tilde n^2}},
\nonumber\\
\tilde\varepsilon_3&=&\frac{q}{\sqrt{-q^2}},
\ear
where
\bear\label{4.3}
\tilde\eta&=&p-\frac{(p_X\cdot p)}{m^2_X}p_X
+\frac{m^2_X(q\cdot p)-(p_X\cdot q)(p_X\cdot p)}
{(p_X\cdot q)^2+Q^2m^2_X}\left[
q-\frac{(p_X\cdot q)}{m^2_X}p_X\right],\nonumber\\
\tilde n^\mu&=&\frac{1}{m_X}\varepsilon^{\mu\nu\rho\sigma}
p_{X\nu}q_\rho p_\sigma=-\frac{m_p}{m_X}n^\mu.\ear
We have as in (\ref{3.10})
\bear\label{4.4}
&&(\tilde\varepsilon_\alpha\cdot \tilde\varepsilon_\beta)=g_{\alpha\beta},
\nonumber\\
&&\det(\tilde\varepsilon_\alpha,\tilde\varepsilon_\beta,\tilde
\varepsilon
_\gamma,\tilde\varepsilon_\delta)=\varepsilon_
{\alpha\beta\gamma\delta}.
\ear
The vectors $\varepsilon_\alpha$ and $\tilde\varepsilon_\alpha$ are
related by a proper, orthochronous Lorentz transformation:
\be\label{4.5}
\tilde\varepsilon_\alpha=\varepsilon_\beta \Lambda^\beta\,_\alpha.\ee
A simple calculation gives for $\Lambda^\beta\,_\alpha$:
\be\label{4.6}
(\Lambda^\beta\,_\alpha)=\left(\begin{array}{cccc}
\gamma_X& -v_X\gamma_X& 0& 0\\
v_X\gamma_X& -\gamma_X& 0& 0\\
0& 0& -1& 0\\
0& 0& 0& 1\end{array} \right),\ee
where
\newpage
\bear\label{4.7}
\gamma_X&=&\left[(p_X\cdot p)(p\cdot q)+Q^2(p_X\cdot p)\right]
\left[(p\cdot q)^2 
+m^2_pQ^2\right]^{-1/2} \left[(p_X\cdot
q)^2+m^2_XQ^2\right]^{-1/2}\nonumber\\ 
&=&\left(1+\frac{4x^2m^2_p}{Q^2}\right)^{-1/2}
\left(1+\frac{4\beta^2t}{Q^2}\right)^{-1/2}
\left(1+\frac{2\beta^2\xi t}{Q^2}\right),\nonumber\\
v_X&=&(1-\gamma_X^{-2})^{1/2}\nonumber\\
&=&\frac{2\beta}{Q}\left[-t(1-\xi)-m^2_p\xi^2
-\frac{t(4m^2_p-t)x^2}{Q^2}\right]^{1/2}
\left(1+\frac{2\beta^2\xi t}{Q^2}\right)^{-1}.\ear
The boost velocity $v_X$ is always small and the
Lorentz factor $\gamma_X$ close to one for the kinematic
range of interest at HERA:
\be\label{4.8}
v_X\approx\frac{2\beta\sqrt{-t}}{Q}\ll 1.\ee

We define now the vectors for right and left circular
polarization of $\gamma^*$ in the system $RX$:
\be\label{4.9}
\tilde\varepsilon_{\pm}=\mp\frac{1}{\sqrt2}
(\tilde\varepsilon_1\pm i\tilde\varepsilon_2)\ee
and the cross sections and interference terms
\be\label{4.10}
\tilde\sigma_{mn}^{(X)}=\frac{4\pi^2\alpha}{K}\tilde\varepsilon_m
^{\mu*}W^{(X)}_{\mu\nu}\tilde\varepsilon_n^\nu \qquad
(m,n=0,\pm).\ee
The linear relation between the $\tilde\sigma_{mn}^{(X)}$
and the $\sigma_{mn}^{(X)}$ is given by a $5\times 5$ matrix
${\cal M}$ for the case which we consider here, i.e.\ where
(\ref{3.26}) holds for $\sigma_{mn}^{(X)}$ and, as is
easily seen, also for $\tilde\sigma_{mn}^{(X)}$:
\be\label{4.11}
\left(\begin{array}{l}
\tilde\sigma_{++}^{(X)}\\
\tilde\sigma_{00}^{(X)}\\
\tilde\sigma_{+-}^{(X)}\\
{\rm Re}\ \tilde\sigma_{+0}^{(X)}\\
{\rm Im}\ \tilde\sigma_{+0}^{(X)}\end{array}
\right)\quad ={\cal M}\quad \left(
\begin{array}{l}
\sigma_{++}^{(X)}\\
\sigma_{00}^{(X)}\\
\sigma_{+-}^{(X)}\\
{\rm Re}\ \sigma_{+0}^{(X)}\\
{\rm Im}\ \sigma_{+0}^{(X)}\end{array}
\right).\ee
The matrix ${\cal M}$ is given in table 3.

\section{Expectations for $\tilde\sigma_{mn}^{(X)}$ in some
models for the reaction $\gamma^*p\to pX$}
\setcounter{equation}{0}
In this section we discuss the expectations one can have for
the quantities $\tilde\sigma_{mn}^{(X)}$
(\ref{4.10}) in various models for the reaction $\gamma^*p\to
pX$.

\subsection{Pomeron exchange models with factorization}
Let us start by discussing the class of models
where the reaction (\ref{1.3}) is viewed as a two-step process:
The original proton emits a Pomeron which collides with
the $\gamma^*$ to give the final hadronic state $X$:
\be\label{5.1}
p(p)\to p(\tilde p)+{\PP}(\Delta),\ee
\be\label{5.2}
\gamma^*(q)+{\PP}(\Delta) \to X(p_X).\ee
By (\ref{5.1}) and (\ref{5.2}) we certainly do \underbar{not}
want to imply that the Pomeron is a particle but only that
the amplitude for (\ref{1.3}) can be written in a
factorizable form:
\be\label{5.3}
A(\gamma^*p\to fp)=A(\gamma^*\PP\to f)\otimes(\PP-
{\rm propagator})\otimes(pp\PP-{\rm vertex\ factor}),
\ee
where $f$ is some final state of $X$ $(f\in X)$.
The proton-proton Pomeron vertex factor is allowed
to have arbitrary Lorentz structure, i.e.\ it can be the sum
of a scalar, a vector, and tensors of arbitrary rank.
Writing out (\ref{5.3}) explicitly we have
then:
\bear\label{5.4}
<f,p(\tilde p)\ {\rm out} |J_\mu(0)|p(p)>&=&{\M}^{f}_\mu\, \bar
u(\tilde p)\Gamma u(p)+{\M}_{\mu,\alpha}^{f}\, \bar u(\tilde p)
\Gamma^\alpha u(p)\nonumber\\
&&
+{\M}^{f}_{\mu,\alpha_1,\alpha_2}\, \bar u(\tilde p)
\Gamma^{\alpha_1,\alpha_2}u(p)+ \ldots \ear
Here $\Gamma,\Gamma^\alpha, \ldots$ stand for ${\PP}$ propagator times
$pp{\PP}$ vertex factor and ${\M}^{f}_\mu$, ${\M}^{f}_{\mu,\alpha},
\ldots$ for the $\gamma^*{\PP}\to f$ amplitudes. We want to consider
all models where the following two hypotheses hold:

(1) The amplitudes ${\M}_\mu^{f},{\M}^{f}_{\mu ,\alpha}, \ldots$
depend only on the momenta $q,\Delta$ and the momenta and
spins of the particles in the final state $f$.

(2) In contrast, the ${\PP}pp$-vertex times ${\PP}$-propagator factors
$\Gamma,\Gamma^\alpha, \ldots$ depend only on the momenta $p,\tilde p$
of the incoming and scattered proton and on the variable $\xi$. The
typical ``effective ${\PP}$-propagator'' in Regge theory has a $\xi$
dependence (cf.~next section) and thus we allow for it in (\ref{5.4}).

With the above we have given a precise definition of what we
call the Pomeron factorization property. In the following
we will show how one can test some aspects of such a
factorizing Pomeron.

An explicit example of a model with factorizing Pomeron is the
Donnachie-Landshoff model \cite{8}, which we will discuss at length in
section 5.2.  Typically in this type of model ${\PP}$-exchange is
built up from a whole series of spin $n = 2,4,6, \ldots$ exchanges.
The coherent sum of these exchanges can give an effective vector-type
coupling of the Pomeron as was exemplified explicitly in a model due
to Van Hove \cite{211} a long time ago (cf.~also chapt.~6.2 of
\cite{11}). We also note the following: The propagator in the usual
field theoretic sense for one of the exchanged particles of definite
spin (e.g. a glueball of spin $n$) depends, of course, only on the
momentum $\Delta$. The corresponding propagator times vertex factor
$\Gamma^{\alpha_1, \ldots, \alpha_n}$ in (\ref{5.4}) has then no
dependence on $q$ and thus no $\xi$ dependence. The latter is
introduced only by the contraction with ${\M}^{f}_{\mu ,\alpha_1,
  \ldots, \alpha_n}$. When the sum over $n$ is performed one gets
Pomeron exchange as an effective vector exchange with a Pomeron
``Propagator'' depending on $\xi$. 

In the rest system of $X$, $RX$ (Fig.~5), the $\gamma^*-{\PP}$
collision is head on. With the criteria I or II
of sect.~2 the set $X$ of final states is
invariant with respect to rotations around the
axis given by the ${\PP}$-momentum ${\bf\Delta}_{RX}$.
We imagine now that we go from the Lorentz
indices $\alpha, \alpha_1, \alpha_2$...
in (\ref{5.4}) to a helicity basis for the
Pomeron with respect to the axis ${\bf\Delta}_{RX}$.
Then (\ref{5.4}) can be rewritten in the form:
\be\label{5.5}
<f,p(\tilde p)\ {\rm out}\ |J_\mu(0)|p(p)>=
{\M}^f_{\mu r}\bar u(\tilde p)\Gamma_r u(p),\ee
where $r=0,\pm1,\pm2,...$ are the helicity indices of ${\PP}$.
(N.B.: The summation convention is always used.)
Inserting (\ref{5.5}) in (\ref{3.6}) and (\ref{4.10}) we get:
\be\label{5.6}
W_{\mu\nu}^{(X)}=S_{\mu r,\nu s}\rho_{s r},\ee
\be\label{5.7}
\tilde\sigma_{mn}^{(X)}=\frac{4\pi^2\alpha}{K}S_{mr,ns}\ \rho_{sr},
\ee
where
\bear\label{5.8}
S_{\mu r, \nu s}&=&\frac{1}{2m_p}\sum_{f\in X}\delta(\Delta+q-p_f)
{\M}^{f*}_{\mu r}\ {\M}^f_{\nu s},\nonumber\\
S_{mr,ns}&=&\tilde\varepsilon^{\mu*}_mS_{\mu r,\nu s}
\tilde\varepsilon^\nu_n,\ear
\be\label{5.9}
\rho_{sr}=\frac{1}{2}\sum_{spins}\,\!\! ^\prime \, \bar u(\tilde p)
\Gamma_s u(p)\, \bar u(p)\bar\Gamma_ru(\tilde p).\ee
Clearly, $(\rho_{sr})$ can be considered as the Pomeron density matrix
in the helicity basis in much the same way as the lepton tensor
$l^{\nu\mu}$ was considered as density matrix of the virtual photon
in sect.~3.

We can state the following:

\underbar{Theorem:} If the Pomeron is described in the system
$RX$ by a density matrix $(\rho_{sr})$
which is diagonal in helicity, then
the interference terms $\tilde\sigma_{mn}^{(X)}, m\not=n$, must
vanish, i.e.
\bear\label{5.10}
\tilde\sigma_{+-}^{(X)}&=&0,\nonumber\\
\tilde\sigma_{+0}^{(X)}&=&0.\ear

The proof of this theorem follows immediately by exploiting
rotational invariance around the common axis of
$\vec q$ and ${\bf\Delta}=-\vec q$ in $RX$ (Fig.~5). The quantity
$S_{mr,ns}$ (\ref{5.8})
can be considered as the generalized ``absorptive part''
 of an amplitude
for a ``reaction''
\be\label{5.11}
\gamma^*{\PP}\to X\to\gamma^*{\PP},\ee
where one does not sum over a complete set of intermediate states,
but only over a set $X$ selected by our criteria I or II of sect.~2.
Now it is important that this set is invariant under rotations
around ${\bf\Delta}_{RX}$. Then the angular momentum around
${\bf\Delta}_{RX}$ for $\gamma^*{\PP}$ in
the initial state of (\ref{5.11}): $s-n$ must be
equal to the one in the final state: $r-m$ (cf.~appendix D for
more details).
Thus we get
\be\label{5.12}
S_{mr,ns}=0\quad{\rm for}\quad m-r\not=n-s.\ee
If we have a diagonal ${\PP}$-density matrix, i.e.\ if
\be\label{5.13}
\rho_{sr}=0\quad{\rm for}\quad
r\not=s\ee
we get from (\ref{5.7}) and (\ref{5.12})
\be\label{5.14}
\tilde\sigma_{mn}^{(X)}=0\quad {\rm for}\quad
m\not= n,\ee
q.e.d.

What can we learn from the theorem (\ref{5.10})?
If $\tilde\sigma_{+-}^{(X)}$ and/or
$\tilde\sigma_{+0}^{(X)}$ are found to be different from zero
in experiment,
we learn in the framework of the ${\PP}$-exchange models
that the Pomeron is not in a diagonal helicity state. This
would imply that the Pomeron cannot have a single helicity,
e.g. helicity zero only. The Pomeron density matrix would
have to contain non-diagonal terms in helicity. More generally,
going beyond the ansatz of a factorizing  Pomeron
(\ref{5.3}),
we would learn that the information on the azimuthal
orientation of the $\vec p-\tilde{\vec p}$ plane in Fig.~5
relative to the polarization directions 1,2 of the photon
$\gamma^*$ must have reached the system $X$. This could have
happened  through the initial state, e.g. by a nondiagonal
helicity density matrix of the exchanged object, as discussed
above, and/or by final state interactions between the remnant
$p(\tilde p)$ and the system $X$. The latter possibility
will be discussed in more detail below in sect.~5.3.

\subsection{The Donnachie-Landshoff Pomeron}

In this section we explore the consequences of the Donnachie-Landshoff
Pomeron model \cite{23,8} for the cross sections and
interference terms $\tilde\sigma_{mn}^{(X)}$  defined in
(\ref{4.10}).
In this model ${\PP}$-exchange is viewed as
exchange of a ``$C=+1$ photon''. The $pp{\PP}$ vertex factor
in Fig.~5 is taken to be proportional to the one for an isoscalar
photon and the ``${\PP}$-propagator'' $P(\xi)$ has the
form \cite{23}:
\be\label{5.15}
P(\xi)=(\xi)^{1-\alpha_{\PP}(t)}\otimes\ {\rm phase\ factor}.
\ee
Here $\alpha_{\PP}(t)$ is the Pomeron-trajectory, determined
experimentally in hadron-hadron scattering as
\bear\label{5.16}
&&\alpha_{\PP}(t)=1+\varepsilon+\alpha't,\nonumber\\
&&\varepsilon=0.08\,\mbox{ to }\, 0.09,\nonumber\\
&&\alpha'=0.25\ {\rm GeV}^{-2}.\ear
The ansatz for $\gamma^*p\to pf$ in this model is then (cf.~(\ref{5.4}))
\bear\label{5.17}
\lefteqn{<f,p(\tilde p)\ {\rm out}\ |J_\mu(0)|p(p)>={\M}^f_{\mu\alpha}
P(\xi) \, i3\beta_q } \nonumber\\
& & \cdot\, \bar u(\tilde
p)\left[F_1(t)\gamma^\alpha-\frac{i}{2m_p}\sigma^ 
{\alpha\beta}\Delta_\beta F_2(t)\right]u(p).\ear
Here
\be\label{5.18}
\beta_q\simeq 1.8\ {\rm GeV}^{-1}\ee
is the quark-Pomeron coupling constant and $F_{1,2}$ are the isoscalar
Dirac and Pauli electromagnetic nucleon form factors, the sum of the
proton and neutron form factors:
\be\label{5.19}
F_j(t)=F_j^p(t)+F_j^n(t)\quad
(j=1,2).\ee

A simple calculation gives now
\be\label{5.20}
W^{(X)}_{\mu\nu}=S_{\mu\alpha,\nu\beta}\rho^{\beta\alpha},\ee
where
\be\label{5.21}
S_{\mu\alpha,\nu\beta}=\frac{1}{2m_p}\sum_{f\in X}\delta(\Delta+q-p_f)
{\M}^{f*}_{\mu\alpha}{\M}^f_{\nu\beta},\ee
\be\label{5.22}
\rho^{\beta\alpha}=9\beta^2_q|P(\xi)|^2\, \tilde\rho^{\beta\alpha},\ee
\bear\label{5.23}
\tilde\rho^{\beta\alpha}&=&\frac{1}{2}\sum_{spins}\,\!\! ^\prime \,
\bar u(\tilde p)\left[F_1\gamma^\beta-\frac{i}{2m_p}\sigma^{\beta\beta'}
\Delta_{\beta'}F_2\right]u(p)\nonumber\\
&&\bar u(p)\left[F_1\gamma^\alpha+\frac{i}{2m_p}\sigma^{\alpha\alpha'}
\Delta_{\alpha'}F_2\right]u(\tilde p)\nonumber\\
&=&2\left(p^\beta-\frac{(p\cdot \Delta)\Delta^\beta}{t}\right)
\left(p^\alpha-\frac{(p\cdot \Delta)\Delta^\alpha}{t}\right)
\tilde G^2(t)\nonumber\\
&&+\frac{t}{2}\left(g^{\beta\alpha}-\frac{\Delta^\beta\Delta^\alpha}{t}
\right)G^2_M(t),\ear
\bear\label{5.24}
G_E(t)&=&F_1(t)+\frac{t}{4m_p^2}F_2(t),\nonumber\\
G_M(t)&=&F_1(t)+F_2(t),\nonumber\\
\tilde G(t)&=&\left(G_E^2(t)-\frac{t}
{4m^2_p}G^2_M(t)\right)^{1/2}\left(1-\frac{t}{4m^2_p}\right)^{-1/2}.\ear
The form factors $G_E$ and $G_M$ are the usual electric and magnetic
isoscalar Sachs form factors. For $t=0$ we have
\bear\label{5.25}
G_E(0)&=&\tilde G(0)=1,\nonumber\\
G_M(0)&=&(\mu_p+\mu_n)/\left(\frac{e}{2m_p}\right)\simeq0.88\ear
where $\mu_p(\mu_n)$ is the magnetic moment of the proton (neutron).
Data for the form factors at values $t<0$ are compiled in \cite{24}.

The problem is now to represent $\tilde\rho^{\beta\alpha}$ in
the helicity basis. This is easily achieved by introducing a vierbein
of ${\PP}$-polarization vectors $\zeta_\lambda$ $(\lambda=0,1,
2,3)$ in analogy to (\ref{3.7})-(\ref{3.9}), but with the
replacements
\be\label{5.26}
p\to p_X,\ q\to\Delta,\ p_X\to p,\ee
which imply, of course, $p^2=m^2_p\to p^2_X=m^2_X$.
This leads to
\bear\label{5.27}
\zeta_0&=&\frac{(p_X\cdot \Delta)\Delta-tp_X}{\sqrt{-t[(p_X\cdot
\Delta)^2-m_X^2t]}},
\nonumber\\
\zeta_1&=&\frac{\zeta_1'}{\sqrt{-{\zeta'}_1^2}},\nonumber\\
\zeta_2&=&\frac{\zeta_2'}{\sqrt{-{\zeta'}_2^2}},\nonumber\\
\zeta_3&=&\frac{\Delta}{\sqrt{-t}},\ear
where
\bear\label{5.28}
\zeta_1'&=&p-\frac{(p_X\cdot p)}{m_X^2}p_X\nonumber\\
&&+\frac{m_X^2(\Delta \cdot p)-(p_X\cdot
\Delta)(p_X\cdot p)}{(p_X\cdot \Delta)^2-m^2_Xt}
\left[\Delta-\frac{(p_X\cdot \Delta)}{m_X^2}p_X\right],\nonumber\\
{\zeta'}^\mu_2&=&\frac{1}{m_X}\varepsilon^{\mu\nu\rho\sigma}p_{X\nu}
\Delta_\rho p_\sigma.\ear
The vectors $\zeta_\lambda$ satisfy
\bear\label{5.29}
(\zeta_{\kappa}\cdot \zeta_\lambda)&=&g_{\kappa\lambda},\nonumber\\
\det(\zeta_0,\zeta_1,\zeta_2,\zeta_3)&=&1.\ear
In the rest system of $X$ we have with ${\bf\widehat\Delta}=
{\bf\Delta}/|{\bf\Delta}|$:
\bear\label{5.30}
\zeta_0|_{RX}&=&\frac{1}{\sqrt{-t}}\left(\begin{array}{c}
|{\bf\Delta}|\\
\Delta^0{\bf\widehat\Delta}\end{array}\right)_{RX},\nonumber\\
\zeta_1|_{RX}&=&\left(\begin{array}{c}
0\\
\frac{\vec p-{\bf\widehat\Delta}({\bf\widehat\Delta}\cdot \vec p)}
{|\vec p-{\bf\widehat\Delta}({\bf\widehat\Delta}\cdot \vec p)|}
\end{array}\right)_{RX},
\nonumber\\
\zeta_2|_{RX}&=&\left(\begin{array}{c}
0\\
\frac{{\bf\Delta}\times\vec p}
{|{\bf\Delta}\times\vec p|}\end{array}\right)_{RX},\nonumber\\
\zeta_3|_{RX}&=&\frac{1}{\sqrt{-t}}\left(\begin{array}{c}
\Delta^0\\
{\bf\Delta}\end{array}\right)_{RX}.\ear
The vectors $\zeta_0$ and $\zeta_3$ correspond to Pomeron
helicity zero and
\be\label{5.31}
\zeta_\pm=\mp\frac{1}{\sqrt2}(\zeta_1\pm i\zeta_2)\ee
to helicity $\pm1$.

The expansion of $\tilde\rho^{\beta\alpha}$ in the helicity
basis is easily obtained. We note first that
\be\label{5.32}
p+\tilde p=a_0\zeta_0+a_1\zeta_1,\ee
where
\bear\label{5.33}
a_0&=&\frac{2\sqrt{-t}}{\xi}\left(1-\frac{1}{2}\xi\right) \left(1+
\frac{4\beta^2t}{Q^2} \right)^{-1/2},\\
\label{5.34}
a_1&=&(a^2_0-4m^2_p+t)^{1/2}\nonumber\\
   &=& \frac{2}{\xi}\left[-t(1-\xi)-m^2_p\xi^2
+\frac{(4m_p^2-t)(-t)\beta^2\xi^2}{Q^2}\right]^{1/2}
\left(1+\frac{4\beta^2t}{Q^2}\right)^{-1/2}.\ear
From (\ref{5.23}) and (\ref{5.32})-(\ref{5.34}) we get
\bear\label{5.35}
\tilde\rho^{\beta\alpha}&=&\frac{1}{2} \left\{ (a_0^2\tilde
G^2+tG^2_M) 
\zeta_0^\beta\zeta_0^{\alpha*} \phantom{\frac{1}{2}} \right. \nonumber\\
&&+\frac{1}{2}(a_1^2\tilde G^2-2tG_M^2)(\zeta_+^\beta
\zeta_+^{\alpha*}+\zeta_-^\beta\zeta_-^{\alpha*})\nonumber\\
&&-\frac{1}{\sqrt2}a_0a_1\tilde G^2(\zeta_0^\beta
\zeta_+^{\alpha*}+\zeta_+^\beta\zeta_0^{\alpha*})\nonumber\\
&&+\frac{1}{\sqrt2}a_0a_1\tilde G^2(\zeta_0^\beta
\zeta_-^{\alpha*}+\zeta_-^\beta\zeta_0^{\alpha*})\nonumber\\
&&\left. -\frac{1}{2}a_1^2\tilde G^2(\zeta_+^\beta
\zeta_-^{\alpha*}+\zeta_-^\beta\zeta_+^{\alpha*}) \right\} \nonumber\\
&=:& \zeta^\beta_s\tilde\rho_{sr}\zeta_r^{\alpha*} \phantom{\frac{1}{2}} \\
\label{5.36}
\rho_{sr}&=&9\beta^2_q|P(\xi)|^2\tilde\rho_{sr}  .\ear
Note that $\zeta_3=\Delta/\sqrt{-t}$ does \underbar{not}
occur in (\ref{5.35}) since the proton-Pomeron coupling is
taken to be proportional to the electromagnetic coupling where
the current is conserved.

We define now in accordance with (\ref{5.8}):
\bear\label{5.37}
S_{\mu r,\nu s}&=&\frac{1}{2m_p}\sum_{f\in X}\delta
(\Delta+q-p_f)({\M}^f_{\mu\alpha}\zeta^\alpha_r)^*({\M}^f_{\nu\beta}
\zeta^\beta_s),\nonumber\\
S_{m r,n s}&=&\frac{1}{2m_p}\sum_{f\in X}\delta
(\Delta+q-p_f)({\M}^f_{\mu\alpha}\tilde
\varepsilon^\mu_m\zeta^\alpha_r)^*({\M}^f_{\nu\beta}
\tilde\varepsilon_n^\nu
\zeta_s^\beta).\ear
From (\ref{5.20}) we get then:
\bear\label{5.38}
W_{\mu\nu}^{(X)}&=&S_{\mu r,\nu s}\rho_{sr},\nonumber\\
\tilde\sigma^{(X)}_{mn}&=&\frac{4\pi^2\alpha}{K}S_{mr,ns}\rho_{sr}.
\ear
These are the explicit forms of (\ref{5.6}),
(\ref{5.7}) in the Donnachie-Landshoff model where the Pomeron helicities
$r,s$ take the values $0,\pm$. Using
(\ref{5.12}) we can write $\tilde\sigma_{mn}^{(X)}$ as follows:
\bear\label{5.39}
\tilde\sigma_{++}^{(X)}&=&\frac{4\pi^2\alpha}{K}\left\{S_{+0,+0}
\rho_{00}+S_{++,++}\rho_{++}+S_{+-,+-}\rho_{--}\right\},\nonumber\\
\tilde\sigma_{00}^{(X)}&=&\frac{4\pi^2\alpha}{K}\left\{S_{00,00}
\rho_{00}+S_{0+,0+}\rho_{++}+S_{0-,0-}\rho_{--}\right\},\nonumber\\
\tilde\sigma_{+-}^{(X)}&=&\frac{4\pi^2\alpha}{K}\left\{S_{++,--}
\rho_{-+}\right\},\nonumber\\
\tilde\sigma_{+0}^{(X)}&=&\frac{4\pi^2\alpha}{K}\left\{S_{++,00}
\rho_{0+}+S_{+0,0-}\rho_{-0}\right\}.\ear

Let us now discuss the behaviour of $\rho_{sr}$ for $\xi\ll1$.
We get from (\ref{5.33})-(\ref{5.36}) in this limit:
\be\label{5.40}
\left( \begin{array}{c} \rho_{00}\\ \rho_{++}=\rho_{--}\\ \rho_{0+}=\rho_{+0}\\
\rho_{0-}=\rho_{-0}\\ \rho_{+-}=\rho_{-+} \end{array} \right) \to
9\beta^2_q|P(\xi)|^2\frac{|t|}{\xi^2}\tilde G^2(t)
\left( \begin{array}{c} 2 \\ 1 \\ -\sqrt2\\ \sqrt2 \\ -1 \end{array}
\right)
\ee
Thus all density matrix elements $\rho_{sr}$ are of the same
order as $\xi\to0$. However, the vector $\zeta_0$ in (\ref{5.30})
contains a factor $1/\sqrt{-t}$ whereas $\zeta_\pm$ (\ref{5.31})
do not. Now the Pomeron is not expected to couple in general to
a conserved current. Especially in inelastic diffraction there is
clear evidence that it does not \cite{8}. Thus we expect
\be\label{5.41}
{\M}^f_{\mu\alpha}\Delta^\alpha=\Delta^0\left({\M}^f_{\mu 0}-{\M}^{fj}
_{\mu}\, \widehat\Delta^j\frac{|{\bf\Delta}|}{\Delta^0}\right)
\Biggr |_{RX}\not=0.\ee
On the other hand, we can write:
\be\label{5.42}
{\M}^f_{\mu\alpha}\zeta^\alpha_0=
\frac{\Delta^0}{\sqrt{-t}}\left({\M}^f_{\mu
0}\frac{|{\bf\Delta}|}{\Delta^0}-{\M}^{fj}_\mu \,
\widehat\Delta^j\right) \Biggr |_{RX}.\ee
We have for $t\to0$
\bear\label{5.43}
&&\frac{|{\bf\Delta}|}{\Delta^0}\Biggr |_{RX}\to 1,\nonumber\\
&&\sqrt{-t}\zeta_0\to\Delta.\ear
We conclude from (\ref{5.41})-(\ref{5.43}) that  
${\M}^f_{\mu\alpha}\zeta_0^\alpha$ will behave as $1/\sqrt{|t|}$
for $t\to0$:
\be\label{5.44}
{\M}^f_{\mu\alpha}\zeta_0^\alpha\to\frac{1}{\sqrt{|t|}}{\M}
^f_{\mu\alpha}\Delta^\alpha.\ee
The matrix elements ${\M}^f_{\mu\alpha}\zeta^\alpha_{\pm}$
on the other hand, have no such factor $1/\sqrt{|t|}$ and are regular
for $t=0$. If we assume that $\sqrt{|t|}\cdot{\M}^f_{\mu\alpha}
\zeta_0^\alpha/\Delta^0_{RX}$ and ${\M}^f_{\mu\alpha}\zeta
^\alpha_{\pm}$ are of the same order of magnitude, we get:
\bear\label{5.45}
S_{\mu 0,\nu 0}&\propto&\frac{(\Delta^0_{RX})^2}{|t|},\nonumber\\
S_{\mu 0,\nu \pm}&\propto&\frac{\Delta^0_{RX}}{\sqrt{|t|}},\nonumber\\
S_{\mu \pm,\nu 0}&\propto&\frac{\Delta^0_{RX}}{\sqrt{|t|}},\nonumber\\
S_{\mu \pm,\nu \pm}&\propto&1,\nonumber\\
S_{\mu \pm,\nu \mp}&\propto&1;\ear
\bear\label{5.46}
W_{\mu\nu}^{(X)}&=&S_{\mu 0,\nu 0}\rho_{00}+\nonumber\\
&&\left\{ \, {\rm longitudinal / transverse\ interference\ terms}
\phantom{\sqrt{|t|}/ \Delta^0_{RX}} 
\right. \nonumber\\ 
&&\left. \:\:\; {\rm in\ Pomeron\ helicity\ of\ relative\ order\ } \sqrt{|t|}
/ \Delta^0_{RX} \right\} + \nonumber\\
&&\left\{ \, {\rm transverse / transverse\ terms} \phantom{\sqrt{|t|}
/ \Delta^0_{RX}} \right. \nonumber\\ 
&&\left. \:\:\; {\rm in\ Pomeron\ helicity\ of\ relative\ order\ }
|t|/(\Delta^0_{RX})^2 \right\},
\ear
where (cf.~table 2):
\be\label{5.47}
\frac{\sqrt{|t|}}{\Delta^0_{RX}}\simeq\frac{2\sqrt{\beta(1-\beta)}
\sqrt{|t|}}{Q}.\ee
The important point in the assumption (\ref{5.45}) is that we use
a \underbar{kinematic} variable of subreaction
(\ref{5.2}) to  compensate the $1/|t|$ and $1/\sqrt{|t|}$ factors
in $S_{\mu 0,\nu 0}$ and $S_{\mu 0, \nu\pm}, S_{\mu\pm,\nu 0}$,
respectively. Using instead of $\Delta^0_{RX}$ the quantities
$q^0_{RX}$ or $Q$ or $m_X$ would not change the result
(\ref{5.46}), (\ref{5.47}) in  an
essential way.

We see by this analysis that the Donnachie-Landshoff Pomeron
couples predominantly with \underbar{longitudinal}
helicity. This implies for the cross sections
\be\label{5.48}
\tilde\sigma_{mn}^{(X)}\simeq\frac{4\pi^2\alpha}{K}S_{m0,n0}
\rho_{00}.\ee
and in particular, with (\ref{5.12})
\be\label{5.49}
\tilde\sigma_{mn}^{(X)}\simeq0\quad{\rm for}\quad
m\not=n.\ee
Thus we come to the following conclusions:

In the Donnachie-Landshoff Pomeron model the interference terms
$\tilde\sigma_{mn}^{(X)}, m\not=n$, should be suppressed, since in
this model the Pomeron helicity is predominantly zero.
We expect from (\ref{5.39}), (\ref{5.40}) and
(\ref{5.45})
\bear\label{5.50}
\frac{\tilde\sigma_{+0}^{(X)}}{\tilde\sigma_{++}^{(X)}+\tilde
\sigma^{(X)}_{00}}&=&{\cal O}\left(\frac{2\sqrt{\beta(1-\beta)}\sqrt{|t|}}
{Q}\right),\nonumber\\
\frac{\tilde\sigma_{+-}^{(X)}}{\tilde\sigma_{++}^{(X)}+\tilde
\sigma^{(X)}_{00}}&=&{\cal O}\left(\frac{4\beta(1-\beta)|t|}
{Q^2}\right).\ear
In this way also (\ref{3.170}) is satisfied for $t=-|t|_{min}$.
Indeed, we are considering here $\xi\ll1$. Since $x\leq\xi$
we have from (\ref{A.7}) $|t|_{min}\approx0$ and from
(\ref{5.50}) that the interference terms $\tilde\sigma_{mn}^{(X)}$
$(m\not=n)$ vanish at $t=0$. Furthermore from (\ref{4.8})
$v_X=0$ at $t=0$ and thus from (\ref{4.11}) and table 3 also
the interference terms $\sigma^{(X)}_{mn}$ $(m\not=n)$
defined in (\ref{3.13})
vanish for $t=0$.

The conclusions reached above for the interference terms in
the Donnachie-Landshoff Pomeron model should be generically
the same in other phenomenological Pomeron exchange models
\cite{21},\cite{26}. But we must leave it to the authors
of these models to give detailed predictions for the interference
terms.

Let us briefly comment on the ``Pomeron structure function''
(cf.~\cite{1} and for a recent review
 \cite{260}) as it presents itself in the light
of our analysis. It is a simple exercise to show that
the totally inclusive $F_2$ structure function of the proton
can be written as follows:
\bear\label{5.51}
F_2^p(x,Q^2)&=&\frac{(p\cdot q)^2}{(p\cdot q)^2+p^2Q^2}\frac{Q^2}
{2(p\cdot q)}\left(\frac{3}{2}\varepsilon_0^\nu\varepsilon_0^{\mu*}
-\frac{1}{2}g^{\nu\mu} \right)
\sum_f(2\pi)^3\delta(p+q-p_f)\nonumber\\
&&\sum_{spins}\,\!\! ^\prime<f|J_\mu(0)|p(p)>^*<f|J_\nu(0)|p(p)>,\ear
where $\varepsilon_0$ is as in (\ref{3.7}). For
\be\label{5.52}
p^2Q^2\ll(p\cdot q)^2\ee
we get
\bear\label{5.53}
F_2^p(x,Q^2)&\simeq&\left(\frac{3}{2}\varepsilon^\nu_0\varepsilon^{\mu*}_0-
\frac{1}{2}g^{\nu\mu}\right) x \sum_f(2\pi)^3\delta(p+q-p_f)\nonumber\\
&&\sum_{spins}\,\!\! ^\prime <f|J_\mu(0)p(p)>^*<f|J_\nu(0)|p(p)>.\ear
For $|t| \to 0$ and with $\xi\ll1$ and (\ref{5.52}) the diffractive
structure function 
$F_2^{D(4)}$ (\ref{3.31}) reads in the
Donnachie-Landshoff Pomeron exchange model, using (\ref{5.37}),
(\ref{5.38}), (\ref{5.46}) and (\ref{5.44}),
\be\label{5.54}
F_2^{D(4)}(x,Q^2,\xi,t)=f(\xi,t)F^{\PP}_2(\beta,Q^2)
\ee
where $f(\xi,t)$ is the conventional ``Pomeron flux factor'':
\be\label{5.55}
f(\xi,t)=\frac{9\beta^2_q}{4\pi^2}|P(\xi)|^2\frac{1}{\xi}
\tilde G^2(t)=\frac{9\beta^2_q}{4\pi^2}\xi^{1-2\alpha_{\PP}(t)}
\tilde G^2(t)\ee
and $F^{\PP}_2$ the ``Pomeron structure function'':
\bear\label{5.56}
F_2^{\PP}(\beta,Q^2)&\approx&\left(\frac{3}{2}\varepsilon_0^\nu
\varepsilon_0^{\mu*}-\frac{1}{2}g^{\nu\mu}\right)
\frac{Q^2}{2(\Delta\cdot q)}\sum_{f\in X}(2\pi)^3\delta(\Delta+q-p_f)
\nonumber\\
&&({\M}^f_{\mu\alpha}\zeta_0^\alpha)^*({\M}^f_{\nu\beta}\zeta
^\beta_0)|t|\nonumber\\
&\approx&\left(\frac{3}{2}\varepsilon^\nu_0\varepsilon_0^{\mu*}-\frac{1}{2}
g^{\nu\mu}\right)\beta\sum_{f\in X}(2\pi)^3\delta(\Delta+q-p_f)\nonumber\\
&&({\M}^f_{\mu\alpha}\Delta^\alpha)^*({\M}^f_{\nu\beta}\Delta^
\beta).\ear
Here we use the replacement (\ref{5.44}) valid for  small $|t|$.
We see from (\ref{3.7}) that $\varepsilon_0$ is unchanged by
the substitution
\be\label{5.57}
p\to \xi p.\ee
This is obvious after setting $m^2_p=p^2$ in (\ref{3.7}).  Thus, in
the approximation $\xi p\approx\Delta$, which is valid for $|t| \to 0$,
the polarization vector $\varepsilon_0$ in (\ref{5.56}) corresponding
to the proton target, can be considered as the appropriate vector for
the ``Pomeron'' target. Comparing (\ref{5.53}) and (\ref{5.56}) we see
then that $F_2^{\PP}$ has indeed the character of a Pomeron $F_2$
structure function since we can make the replacements
\bear\label{5.58}
{\rm proton}\quad&\longrightarrow&\quad{\rm Pomeron};\nonumber\\
p^\mu\quad&\longrightarrow&\quad\Delta^\mu,\nonumber\\
x\quad&\longrightarrow&\quad\beta,\nonumber\\
<f|J_\mu(0)|p(p)>\ &\longrightarrow&\ {\M}^f_{\mu\alpha}
\zeta_0^\alpha\sqrt{|t|}\approx{\M}^f_{\mu\alpha}\Delta^\alpha\ear
where we use also (\ref{5.44}).
However, we see that $F_2^{\PP}$ is the ``structure function''
of a Pomeron with \underbar{longitudinal} helicity, with polarization
vector $\sqrt{|t|}\zeta_0$. This is \underbar{not} a
\underbar{normalized} polarization vector,
and already from this observation we should not expect
any kind of momentum sum rule for the Pomeron structure function,
i.e.\ its parton densities, to hold.

\subsection{Soft colour interaction and ``background field'' models}
In this section we will first make some remarks on soft
colour interaction models \cite{13}, \cite{14}. Then we will
discuss possible consequences of a picture where the QCD vacuum
``background fields'' play an important role \cite{27}-\cite{30}.

In \cite{13},\cite{14} the reaction (\ref{1.3}) is viewed as proceeding
in 3 steps:

(1) emission of a gluon by the original proton, leaving it as
a coloured remnant $\tilde p_c$
\be\label{5.59}
p(p)\longrightarrow G(\Delta)+\tilde p_c(\tilde p),\ee

(2) $\gamma^*-G$ fusion to give a coloured state $X_c$
\be\label{5.60}
\gamma^*(q)+G(\Delta)\to X_c(p_X),\ee

(3) soft colour exchange where momenta are not changed:
\be\label{5.61}
\tilde p_c(\tilde p)+X_c(p_X)\to p(\tilde p)+X(p_X).\ee

The gluon emission in (\ref{5.59}) should be governed by
a vertex factor with one vector index $\alpha$, quite similar
to the Pomeron emission in (\ref{5.17}). Thus, the first guess
would be that everything should be quite similar for the
soft colour interaction models and the Pomeron exchange models.
However, this is not necessarily so. To see this, we will consider
replacing the gluon in (\ref{5.59}) and (\ref{5.60}) by a
hypothetical ``isoscalar photon'' $\tilde\gamma$ which couples
to a \underbar{conserved} current. Then (\ref{5.17}) is still valid,
but with $P(\xi)i3\beta_q$ replaced by a propagator factor for
$\tilde\gamma$:
\be\label{5.62}
P(\xi)\, i3\beta_q\to P_{\tilde\gamma}(t).\ee
With this all the analysis from (\ref{5.19}) to (\ref{5.40})
remains true. However, instead of (\ref{5.41}) we find now
\be\label{5.63}
{\M}^f_{\mu\alpha}\Delta^\alpha=0.\ee
We have
\be\label{5.64}
\zeta_0|_{RX}=\frac{1}{\sqrt{|t|}}\Delta+\frac
{\sqrt{|t|}}{|{\bf\Delta}_{RX}|+\Delta^0_{RX}}
{1 \choose -{\bf\widehat\Delta}}_{RX}.\ee
Together with (\ref{5.63}) this leads to a \underbar{suppression}
of longitudinal $\tilde\gamma$-helicity contributions for $|t|\to0$.
Instead of (\ref{5.45}) we find here:
\bear\label{5.65}
&&S_{\mu 0,\nu 0}\propto\frac{|t|}{(\Delta^0_{RX})^2},\nonumber\\
&&S_{\mu 0,\nu \pm}, S_{\mu \pm,\nu
0}\propto\frac{\sqrt{|t|}}{\Delta^0_{RX}},\nonumber\\
&&S_{\mu \pm,\nu \pm}, S_{\mu\pm,\nu \mp}\propto1.\ear
Thus, with exchange of the hypothetical $\tilde\gamma$
``photon'', \underbar{transverse} $\tilde\gamma$-helicities
dominate for small $|t|$ and the interference term
$\tilde\sigma_{+-}$ could be large.

We note that the discussions leading to (\ref{5.45}) and (\ref{5.65}),
respectively, are completely analogous to the discussions for PCAC
and CVC tests in inelastic neutrino-nucleon scattering \cite{261}.

In the soft colour exchange models the gluon in (\ref{5.59}),
(\ref{5.60}) should behave like our $\tilde\gamma$ above. However,
the last step (\ref{5.61}) could, a priori, change the situation
and invalidate (\ref{5.63}). We must again leave it to the
authors of \cite{13},\cite{14} to give their predictions for the
interference terms $\tilde\sigma_{mn}^{(X)}\ (m\not=n)$.

In the remainder of this section we will discuss the idea that the QCD
``vacuum background fields'' may play an important role in
(\ref{1.3}). We take the reaction (\ref{1.3}) to proceed as in
\cite{13},\cite{14} in the three steps (\ref{5.59})-(\ref{5.61}), but
in the spirit of \cite{27}-\cite{30} we assume them to take place in a
fluctuating vacuum background field.  For the hard reaction
(\ref{1.3}) at high $Q^2$ we take the vacuum fields as ``frozen''
(Fig.~6). It is clear that the interaction with the background field
will in general invalidate angular momentum conservation arguments for
the $\gamma^* G$ collision. The chromomagnetic Lorentz force will, for
instance, lead to a correlated deflection of the coloured objects $q,
\bar q, G$ and $\tilde p_c$. The background field may lead both to
colour transfer between $X_c$ and $\tilde p_c$ and to angular momentum
transfer. It may also lead to spin flips of $q$ and $\bar q$ in a way
similar to the one considered for the Drell-Yan process in \cite{28}.
In such a scenario we can expect for the diffractive $\gamma^* p$
reaction \underbar{large} interference terms $\tilde\sigma_{mn}^{(X)}$
$(m\not=n)$ which are in fact quantities quite analogous to the
structure function $\nu$ in the Drell-Yan case (cf.~\cite{28}).
Experiments for the Drell-Yan reaction $\pi^- N\to\gamma^* X$
\cite{31} find rather large values of $\nu$ in contrast to the
expectation from the QCD-improved parton model \cite{32}.  Returning
to our interference terms $\tilde\sigma_{mn}^{(X)}$ $(m\not= n)$, we
note that, of course, the boundary conditions (\ref{3.170}) for $t \to
- |t_{\rm min}|$ must be respected. But guided by the Drell-Yan case we
could expect here instead of (\ref{5.50}) a behaviour:
\bear\label{5.66}
&&\frac{\tilde\sigma_{+0}^{(X)}}{\tilde\sigma_{++}^{(X)}+
\tilde\sigma_{00}^{(X)}}={\cal O}(\frac{\sqrt{|t|}}{M_c}),\nonumber\\
&&\frac{\tilde\sigma_{+-}^{(X)}}{\tilde\sigma_{++}^{(X)}+
\tilde\sigma_{00}^{(X)}}={\cal O}(\frac{|t|}{M_c^2}),\ear
where $M_c$ is a hadronic scale, $M_c\simeq 1,0-1,5$ GeV say.
Clearly, experiments should be able to distinguish (\ref{5.50})
and (\ref{5.66}). Especially in inelastic diffraction,
where the proton breaks up and where the $|t|$-dependences of
the cross sections are expected to be much less steep than
in elastic diffraction, measurements over a large $t$-range
may be feasible.

\section{Odderon contributions in $\gamma^*p\to\tilde pX$}
\setcounter{equation}{0}
Let us finally make some remarks on possible ``Odderon''
contributions in the reaction (\ref{1.3}). The Odderon,
the $C=-1$ partner of the Pomeron, was discussed in the
framework of Regge theory in \cite{33}. In the framework
of soft reactions in QCD developed in \cite{10},
\cite{11} there should be an Odderon. In essence, the argument
is as follows. There is no small coupling constant in the
soft region. Thus, the exchange of two nonperturbative gluons
between quarks, which gives the simplest contribution
to the Pomeron, should be comparable to the exchange of three
nonperturbative gluons which can lead to an Odderon contribution \cite{390}.
In perturbation theory, again one finds Odderon contributions
\cite{34}. Phenomenology of soft hadronic reactions, especially
$pp$ and $p\bar p$ scattering finds, on the other hand, no
significant contribution from the Odderon at small $|t|$.
A possible solution to this puzzle is proposed in \cite{35}:
The Odderon coupling to quarks is intrinsically large, but
its coupling to protons is suppressed. This happens if the proton
has a quark-diquark structure.

In our present article we just want to point out again
the possibility of looking for Odderon contributions in diffractive
reactions initiated by a virtual photon (cf.~\cite{36}).
Ideal reactions are ones where $X$ in (\ref{1.3}) is an exclusive
state with charge conjugation quantum number $C=+1$.
\be\label{6.1}
\gamma^*+p\to\tilde p+X_{excl.}(C=+1),\ee
with, for instance,
\be\label{6.2}
X_{excl.}=\pi^0,\eta,\eta',\eta_c,f_2(1270).\ee
Estimates for pseudoscalar production were given in \cite{36}.
Here we will only make three more remarks:

The Odderon exchange for a reaction (\ref{6.1}) where the
virtual photon $\gamma^*$ has high $Q^2$ may be quite strong,
even if in soft hadronic reactions the Odderon is practically absent.

If the Odderon suppression is due to the structure of the proton
wave function as advocated in \cite{35} we can expect that there
is no longer such a suppression for inelastic diffraction, i.e.\ for
the proton remnant $\tilde p$ different from a single proton.

In the scenario of QCD vacuum background fields, discussed at
the end of section 5.3, we should expect large Odderon
contributions and thus large cross sections for the reactions
(\ref{6.1}), (\ref{6.2}). Suppose, for instance, that the
background field easily makes spin flips of $q$ and $\bar q$ in Fig.
6, such that their spin orientation is random  after some time.
Then we should expect from the number of available spin states:
\be\label{6.3}
\frac{\sigma(\gamma^*p\to\tilde p\pi^0)}{\sigma(\gamma^*p
\to\tilde p\rho^0)}\simeq\frac{1}{3}\ee
If, on the other hand, a chromomagnetic Sokolov-Ternov effect,
i.e.\ a build-up of transverse polarizations of $q$ and $\bar q$ as
they travel through the chromomagnetic vacuum fields
(cf.~\cite{27}-\cite{30}) is at work, we can expect to find the
$q\bar q$ pair with oppositely oriented spins which would lead
us even to the expectation
\be\label{6.4}
\frac{\sigma(\gamma^*p\to\tilde p\pi^0)}{\sigma(\gamma^*p
\to \tilde p\rho^0)}\simeq1.\ee
Similarly we could expect
\be\label{6.5}
\frac{\sigma(\gamma^*p\to\tilde p\eta)}{\sigma(\gamma^*p
\to \tilde p\rho^0)}\simeq\frac{\cos^2\vartheta}{9}\
{\rm or}\ \frac{\cos^2\vartheta}{3}\ee
for the above two scenarios. Here $\vartheta\simeq-20^\circ$
is the flavour octet-singlet mixing angle for the pseudoscalar
mesons \cite{pdg}.

We arrive at (\ref{6.3})-(\ref{6.5}) as follows. In the
constituent quark picture the flavour wave functions
for $\pi^0,\rho^0$ and $\eta$ are
\bear\label{6.6}
\pi^0 &\sim &\frac{1}{\sqrt2}(u\bar u-d\bar d),\nonumber\\
\rho^0 &\sim &\frac{1}{\sqrt2}(u\bar u-d\bar d),\nonumber\\
\eta &\sim &\cos\vartheta\, \eta_8+\sin\vartheta\, \eta_1\nonumber\\
&=&\cos\vartheta\frac{1}{\sqrt6}(u\bar u+d\bar d-2s\bar s)
+\sin\vartheta\frac{1}{\sqrt3}(u\bar u+d\bar d+s\bar s)\nonumber\\
\eta' &\sim& -\sin\vartheta \, \eta_8+\cos\vartheta\, \eta_1.\ear
From the diagram of Fig.~6 we estimate the amplitudes for $\gamma^*
+p\to\tilde p$ + (meson) to be proportional to the sum of the
products of the quark charges times the corresponding factor
in front of $u\bar u,d\bar d$ and $s\bar s$, respectively,
in the wave functions. In this way we get the following flavour
factors for the amplitudes:
\bear\label{6.7}
&&A(\gamma^*p\to\tilde p\pi^0)\propto\frac{2}{3}\frac{1}{\sqrt2}
+\frac{1}{3}\frac{1}{\sqrt2}=\frac{1}{\sqrt2},\nonumber\\
&&A(\gamma^*p\to\tilde p\rho^0)\propto\frac{2}{3}\frac{1}{\sqrt2}
+\frac{1}{3}\frac{1}{\sqrt2}=\frac{1}{\sqrt2},\\
&&A(\gamma^*p\to\tilde p\eta)\propto\cos\vartheta\frac{1}{\sqrt 6}
\left(\frac{2}{3}-\frac{1}{3}
+\frac{2}{3}\right)+\sin\vartheta\frac{1}{\sqrt3}
\left(\frac{2}{3}-\frac{1}{3}-\frac{1}{3}\right)=\frac{\cos\vartheta}{\sqrt6}.
\nonumber
\ear
Squaring and multiplying with the spin factors 1/3 or 1 we
arrive at the estimates (\ref{6.3})-(\ref{6.5}). In
the same approximation we expect
\be\label{6.8}
\frac{\sigma(\gamma^*p\to\tilde p\eta')}{\sigma(\gamma^*p
\to \tilde p\eta)}\simeq\tan^2\vartheta\simeq 0.13 \, . \ee

Of course, the final states in reactions (\ref{6.1}), (\ref{6.2}) can
also be reached by the electromagnetic process where the Odderon is
replaced by a virtual photon. The amplitudes for the Odderon and
photon exchanges have to be added and thus there is interference which
can be constructive or destructive, depending on the sign of the
Odderon couplings. For further details we refer to \cite{36}, where,
however, the scale on the abscissa of Fig.~3 is incorrect. An updated
version of this paper is in preparation. The rate estimates given
above apply to the case where the Odderon exchange dominates over
photon exchange.

With this we conclude our brief discussion of ``Odderon physics''
in $\gamma^*p$ collisions.

\section{Conclusions}
In this article we have first advocated to analyse the production of
rapidity gap events at HERA, i.e.\ the reaction $\gamma^*p\to
\tilde pX$ in a way which exploits the kinematic similarity
with one pion electroproduction, $\gamma^*p\to p\pi^0$.
We introduced the azimuthal angle $\varphi$
between the leptonic and a hadronic plane
in the rest system of the original proton and showed
how $\varphi$ is related to quantities directly measurable in the
HERA system. The dependence of the differential cross section
on $\varphi$ was made completely explicit in sect.~3, using
the fact that the photon $\gamma^*$ exchanged between the lepton
and the hadrons in Fig.~1 is a vector particle. We defined cross
sections and interference terms $\sigma_{mn}^{(X)}$ for the
absorption of photons of helicities $m,n=0,+,-$ in the proton
rest frame and showed how these can be isolated using the
$\varphi$ and $\varepsilon$-dependence of the cross section.
The main result there is  (\ref{3.28}), where
the $\varphi$ and $\varepsilon$ dependence is completely explicit
and the $\sigma^{(X)}_{mn}$ are functions of the other kinematic
variables $x,Q^2,\xi,t$ only.
 In sect.~4 the reaction $\gamma^*p\to
\tilde pX$ was considered in the $X$ rest system, where we
could again define cross sections and interference terms
$\tilde\sigma_{mn}^{(X)}$. The usual structure function
for diffractive events, $F_2^{D(4)}$ is related to
$\sigma_{++}^{(X)}+\sigma_{00}^{(X)}$. The interference
terms $\sigma_{mn}^{(X)}(m\not=n)$ are additional
measurable quantities which can be used to test
various models for diffractive events. In sect.~5 we investigated
the class of models where a factorizing Pomeron is assumed to
be responsible for the diffractive events. We have shown that
in such models the interference terms $\tilde\sigma^{(X)}_{mn}
(m\not=n)$ must vanish if the Pomeron is described by a density
matrix which is diagonal in helicity. Thus, a nonzero
$\tilde\sigma^{(X)}_{+0}$ and/or  $\tilde\sigma^{(X)}_{+-}$
would rule out all models where the Pomeron has only a single
helicity, for instance helicity zero. We have then analysed
in detail the Donnachie-Landshoff Pomeron model. We have
shown that there the Pomeron has predominantly helicity
zero with the interference terms  $\tilde\sigma^{(X)}_{+0}$
and  $\tilde\sigma^{(X)}_{+-}$ being suppressed by
$\sqrt{|t|}/Q$ and $|t|/Q^2$, respectively.
We argued that in soft colour interaction models, especially
if we considered possible effects of QCD vacuum background fields,
the interference terms $\tilde\sigma^{(X)}_{+0}$ and/or
 $\tilde\sigma^{(X)}_{+-}$ could be expected to be sizable.
Finally, in sect.~6, we discussed possible signals for an
Odderon contribution in diffractive scattering, $\gamma^*p\to
\tilde p X_{excl.}$ and speculated that such contributions
could be quite large.

We hope to have given in this paper some suggestions for
experimentalists what to look for in diffractive scattering
at HERA in order to further elucidate the underlying
dynamical mechanisms.
\bigskip

\section*{Acknowledgements}
The authors have profited from conversations with H. Abramowicz, J.
Bartels, W.~Buchm\"uller, H. G. Dosch, F. Eisele, L. Frankfurt, G.
Ingelman, H. Jung, K. Meier, H. J. Pirner, M. Rueter and S. Tapprogge. The ARC grant
313-ARC-VIII-VO/scu is gratefully acknowledged which made the mutual
visits of the Cambridge and Heidelberg groups possible. M.D. and
P.V.L. also acknowledge the support of PPARC and of the EU Programme 
``Human Capital and Mobility'', Network ``Physics at High Energy
Colliders'', Contract CHRX-CT93-0357 (DG 12 COMA), and Contract
ERBCHBI-CT94-1342.

\section*{Appendix A}
\renewcommand{\theequation}{A.\arabic{equation}}
\setcounter{equation}{0}
In this appendix we will discuss in which kinematic situations
$n^2=0$ occurs where $n$ is defined in (\ref{2.5}).
For $n^2=0$ the angle $\varphi$ is undefined (cf.
(\ref{2.7})). The discussion is best done in the
$\gamma^*$-proton c.m.\ system (hcm), which is also the
c.m.\ system of the final state hadrons.  We have from (\ref{2.5})
\be\label{A.1}
n_{hcm}=\frac{W}{m_p}\left(\begin{array}{c}
0\\
\vec q\times\vec p_X\end{array}\right)_{hcm}.\ee
This shows that $n^2$ vanishes if and only if $\vec q_{hcm}$
and ${\vec p_{X}|}_{hcm}$ are collinear. This is the case for $t=
-|t|_{\rm min}$ and $t=-|t|_{\rm max}$. Here $|t|_{\rm min}$ $(|t|
_{\rm max})$ is the minimal (maximal) value of $|t|$ at fixed
$W^2,Q^2,\tilde m^2,m^2_X$ which is reached  for $\tilde{\vec p}
_{hcm}=-{\vec p_{X}|}_{hcm}$ parallel (antiparallel) to $\vec p_{hcm}
=-\vec q_{hcm}$.
For diffractive scattering the case $t=-|t|_{\rm max}$ (back scattering)
 is expected to correspond to very small cross section, and thus
is of no practical interest.
A simple calculation gives
\be\label{A.2}
|t|_{\rm min\atop \rm max}=(2p^0\tilde p^0\mp 2|\vec p|
|\tilde {\vec p}|-m^2_p-\tilde m^2)_{hcm},\ee
\bear\label{A.3}|t|_{\rm min} &=&\left\{
\tilde m^2\left(\frac{2p^0}{\tilde p^0+|\tilde{\vec  p}|}-1
\right)\right. +m^2_p\left(\frac{2\tilde p^0}{p^0+|\vec p|}-1\right)
\nonumber\\
&&\left.-2\frac{m^2_p\tilde m^2}{(p^0+|\vec p|)(\tilde p^0+
|\tilde{\vec p}|)}\right\}_{hcm},\ear
\bear\label{A.4} p^0_{hcm}&=&\frac{Q^2}{2Wx}\left(1+\frac{
2m^2_px}{Q^2}\right),\nonumber\\
|\vec p|_{hcm}&=&\frac{Q^2}{2Wx}\left(1+\frac{4m^2_px^2}{Q^2}
\right)^{1/2},\ear
\bear\label{A.5}
\tilde p^0_{hcm}&=&\frac{1}{2W}\frac{(1-x)Q^2-x m^2_X}{x}
\left[1+\frac{x(m^2_p+\tilde m^2)}{(1-x)Q^2-xm^2_X}
\right],\nonumber\\
|\tilde{\vec p}|_{hcm}&=&\frac{1}{2W}\frac{(1-x)Q^2-x m^2_X}{x}
\left\{1+\frac{2x(m^2_p+\tilde m^2)}{(1-x)Q^2-x m^2_X}\right.
\nonumber\\
&&-\frac{4x(1-x)Q^2\tilde m^2}{[(1-x)Q^2-xm^2_X]^2}
\left.+\frac{x^2(\tilde m^2-m^2_p)^2}{
[(1-x)Q^2-xm^2_X]^2}\right\}^{1/2}.\ear
For $m_X^2$ of order $Q^2$ (cf.~table 2) and $Q^2\gg m^2_p,
\tilde m^2$ we obtain
\bear\label{A.6}
|t|_{\rm min}&=&x\frac{Q^2+m^2_X}{Q^2}\left[\tilde
m^2\frac{Q^2}{Q^2-x(Q^2+m^2_X)}
-m^2_p\right]\nonumber\\
&&+{\cal O}\left(\frac{m^4_p}{Q^2},\ \frac{\tilde m^4}{Q^2},\
\frac{m^2_p\tilde m^2}{Q^2}\right).\ear
For the case of the proton remnant being again a single
proton we get with $\tilde m^2=m_p^2$:
\be\label{A.7}
|t|_{\rm min}=x^2\frac{(Q^2+m^2_X)^2}{Q^2[Q^2-x(Q^2+m^2_X)]}m^2_p+
{\cal O}\left(\frac{m^4_p}{Q^2}\right).\ee

\section*{Appendix B}
\renewcommand{\theequation}{B.\arabic{equation}}
\setcounter{equation}{0}
In this appendix we discuss the more general reaction ($n=1,2,...$;
$\tilde n=1,2,...$)
\bear\label{B.1}
e^\mp(k)+p(p)\to e^\mp(k')&+&a_1(p_1)+...+a_n(p_n)\nonumber\\
&+&\tilde a_1(\tilde p_1)+...+\tilde a_{\tilde n}(\tilde p
_{\tilde n}),\ear
where the proton remnant is a group of particles $\tilde a_1,...,
\tilde a_{\tilde n}$ with momenta $\tilde p_1,...,\tilde p_{\tilde n}$
and
\be\label{B.2}
\sum^{\tilde n}_{j=1}\tilde p_j=\tilde p.\ee
In order to make the dependence of the cross section for
the reaction (\ref{B.1}) on $\varphi$ again completely
explicit, one can use the criteria III or IV given in
sect.~2 to define the hadronic system $\tilde p$. When
the proton remnant is no longer a single proton, one needs
an additional kinematical variable, e.g. $\tilde m^2\equiv \tilde p^2$
in addition to $\vec k'$ and $\tilde{\vec p}$ (cf.~(\ref{3.4}))
to describe the process (\ref{1.1}) completely. Now, the cross
section for the reaction (\ref{1.1}) reads
\bear\label{B.3}
d\sigma(e^\mp p\to e^\mp\tilde p X)=\frac{4m_p}{s-m^2_p}
\left(\frac{\alpha}{Q^2}\right)^2l_{\nu\mu}W^{(X,\, \tilde p)\mu
\nu}\frac{d^3k'}{{k'}^0}\frac{d^3\tilde p}{\tilde p^0}d\tilde m^2\ear
where
\bear\label{B.4}
W^{(X, \, \tilde p)\mu\nu}&=&\frac{1}{8\pi m_p}\sum_n\sum_{\tilde n}
\int_{p_1}...\int_{p_n}\int_{\tilde p_1}...\int_{\tilde p_{\tilde n}}
\prod^n_{i=1}\frac{d^3p_i}{(2\pi)^32p_i^0}\nonumber\\
&&\prod^{\tilde n}_{j=1}\frac{d^3\tilde p_j}{(2\pi)^32{\tilde p}_j^0}
\sum_{spins}\,\!\! ^\prime <p(p)|J^\mu (0)|a_i(p_i)\tilde a_j
(\tilde p_j)\ \rm{out}\ >\nonumber\\
&&<a_i(p_i)\tilde a_j(\tilde p_j)\ \rm{out}\ |J^\nu (0)|p(p)>\delta
(\tilde p-\sum^{\tilde n}_{j=1}\tilde p_j)\nonumber\\
&&\chi_n(p_1,...,p_n,p,q)\tilde\chi_{\tilde n}(\tilde p_1,...,
\tilde p_{\tilde n},p,q)\nonumber\\
&&(2\pi)^4\delta(p+q-\tilde p-\sum^n_{i=1}p_i).\ear
For criterion III $\chi_n$ is given by (\ref{3.171}),
$\chi_n^{(III)}=\chi_n^{(I)}$ and
\be\label{B.5}
\tilde \chi_n^{(III)}(\tilde p_1,...,\tilde p_{\tilde n},p,q)=
\Theta(\tilde m_{cut}^2-(\tilde p_1+...+\tilde p_{\tilde n})^2).\ee
For criterion IV we have to set $\chi_n^{(IV)}=\chi_n^{(II)}$ from
(\ref{3.18}) and
\be\label{B.6}
\tilde\chi^{(IV)}_n(\tilde p_1,...,\tilde p_n,p,q)=\prod
^{\tilde n}_{j=1}\Theta(\tilde M_{cut}^2-\tilde p \cdot \tilde p_j).\ee
With the definition (cf.~(\ref{3.13})):
\be\label{B.7}
\sigma_{mn}^{(X, \, \tilde p)}:=\frac{4\pi^2\alpha}{K}
\varepsilon^{*\mu}_mW^{(X,\, \tilde p)}_{\mu\nu}\varepsilon_n^\nu\ee
we find again that the relations (\ref{3.14}), (\ref{3.15})
and (\ref{3.25}) hold. In analogy to (\ref{3.28}) we get then
\bear\label{B.8}
\frac{d^7\sigma(ep\to e\tilde pX)}{dx\, dQ^2\, d\xi\, dt\, 
d\varphi\, d\tilde m^2 d\tilde\psi}
&=&\frac{1}{2\pi}\tilde\Gamma\{\sigma_{++}^{(X,\,\tilde p)}+\varepsilon
\sigma_{00}^{(X,\, \tilde p)}\nonumber\\
&&-\varepsilon\cos2\varphi\sigma_{+-}^{(X,\, \tilde p)}\nonumber\\
&&-2\sqrt{\varepsilon(1+\varepsilon)}\cos\varphi\ {\rm Re}\
\sigma_{+0}^{(X,\, \tilde p)}\nonumber\\
&&+2r_L\sqrt{\varepsilon (1-\varepsilon)}\sin\varphi\
{\rm Im}\ \sigma_{+0}^{(X,\, \tilde p)}\}.\ear
The cross sections and interference terms $\sigma_{mn}^{(X,\, \tilde p)}$
for definite $\gamma^*$ helicities now depend on 5 kinematical variables
\be\label{A.9}
\sigma_{mn}^{(X,\, \tilde p)}=\sigma_{mn}^{(X,\, \tilde p)}(x,Q^2,
\xi,t,\tilde m^2).\ee
The analyses presented in sections 4, 5, and 6 for the case of
elastic diffraction ($\tilde p$= single proton), are - with obvious
replacements - valid also for the more general case of inelastic
diffraction considered in this appendix.

\section*{Appendix C}
\renewcommand{\theequation}{C.\arabic{equation}}
\setcounter{equation}{0}
 In (\ref{3.15}) we asserted that the matrix $(\sigma_{mn}^{(X)})$
must be positive semi-definite. The condition for this is
that for every complex vector $(c_m)$
\be\label{C.1}
\sum_{m,m'}c_m^*\sigma_{mm'}^{(X)}c_{m'}\geq0.\ee
Indeed, inserting the definition of  $\sigma_{mm'}^{(X)}$ from
(\ref{3.13}), (\ref{3.6}) we get
\bear\label{C.2}
&&\sum_{m,m'}c_m^*\sigma_{mm'}^{(X)}c_{m'}=
\frac{4\pi^2\alpha}{K}\sum_{m,m'}c_m^*\varepsilon_m^{*\mu}
W_{\mu\nu}^{(X)}\varepsilon^\nu_{m'}c_{m'}\nonumber\\
&&=\frac{4\pi^2\alpha}{K}\frac{1}{4m_p}\sum_n\int_{p_1}
...\int_{p_n}\prod^n_{i=1}\frac{d^3p_i}{(2\pi)^32p_i^0}\nonumber\\
&&\chi_n(p_1,...,p_n,p,q,\tilde p)
\delta(p+q-\tilde p-\sum^n_{i=1}p_i)\nonumber\\
&&\sum_{spins}\,\!\! ^\prime |\sum_{m'}<a_i(p_i)p(\tilde p)\ out\ |
J_\nu(0)|p(p)>\varepsilon_{m'}^\nu c_{m'}|^2\geq0.\ear

Under the assumption that (\ref{3.26}) holds, the matrix $(\sigma_{mn}
^{(X)})$ reads
\be\label{C.3}
\underline{\sigma}=\left(\begin{array}{ccc}
\sigma_{00}^{(X)}&\sigma_{0+}^{(X)}&-\sigma_{0+}^{(X)}\\
\sigma_{+0}^{(X)}&\sigma_{++}^{(X)}&\sigma_{+-}^{(X)}\\
-\sigma_{+0}^{(X)}&\sigma_{+-}^{(X)}&\sigma_{++}^{(X)}\end{array}
\right)\ee
with eigenvalues
\bear\label{C.4}
\lambda_1&=&\sigma_{++}^{(X)}+\sigma_{+-}^{(X)}\nonumber\\
\lambda_{2,3}&=&\frac{1}{2}\Bigl[
\sigma_{00}^{(X)}+\sigma_{++}^{(X)}-\sigma_{+-}^{(X)}\nonumber\\
&&\pm\sqrt{(\sigma_{00}^{(X)}+\sigma_{++}^{(X)}-\sigma_{+-}^{(X)})^2
+4(\sigma^{(X)}_{+-}
\sigma_{00}^{(X)}+2|\sigma_{+0}^{(X)}|^2-\sigma_{++}^{(X)}
\sigma_{00}^{(X)})}\Bigr].\ear
The necessary and sufficient condition for the matrix $(\sigma_{mn}^{(X)})$
to be positive semi-definite is
\be\label{C.5}
\lambda_i\geq0,\quad i=1,2,3.\ee
Condition (\ref{C.5}) is equivalent to
\bear\label{C.6}
&&\sigma_{00}^{(X)}\ge 0,\nonumber\\
&&\sigma_{++}^{(X)}\pm\sigma_{+-}^{(X)}\ge 0,\nonumber\\
&&\sigma_{00}^{(X)}(\sigma_{++}^{(X)}-\sigma_{+-}^{(X)})\ge
2|\sigma_{+0}^{(X)}|^2.
\ear
These could be used in an experimental analysis in the following way. Suppose 
one has measured the $\varphi$-integrated cross section (cf.~(\ref{3.30})):
\bear\label{C.7}
\sigma_{\varepsilon}=\sigma_{++}^{(X)}+\varepsilon\sigma_{00}^{(X)}.
\ear
Typically one has only a small range in $\varepsilon$ available and a
separation of $\sigma_{++}^{(X)}$ and $\sigma_{00}^{(X)}$ is very difficult.
Let us for simplicity assume that just one value of $\varepsilon$ is available, 
but that in addition to $\sigma_{\varepsilon}$ one has measured the
interference terms $\sigma_{+-}^{(X)}$ and ${\rm Re}\sigma_{+0}^{(X)}$ (cf.
(\ref{3.290})). Then the allowed range for $\sigma_{00}^{(X)}$ and
$\sigma_{++}^{(X)}$ in the $\sigma_{++}^{(X)}$ - $\sigma_{00}^{(X)}$ plane can 
be constructed as shown in Figs. 7(a,b). In particular we see that
${\rm Re}\sigma_{+0}^{(X)}\neq 0$ implies,
of course, $\sigma_{00}^{(X)}\neq 0$.

\section*{Appendix D}
\renewcommand{\theequation}{D.\arabic{equation}}
\setcounter{equation}{0}

Here we give the full details of the argument leading to (\ref{5.12}).
Consider the amplitude for $\gamma^*$ with helicity $n$ and $\PP$
with helicity $s$ leading to a final state $f\in X$ (cf.~(\ref{5.5}) -
(\ref{5.8})):
\be\label{D.1}
{\cal M}^f_{ns}={\cal M}^f_{\mu s}\tilde\varepsilon^\mu_n,\ee
where we always work in the system $RX$. Under a rotation $R(\chi)$
by an arbitrary angle $\chi$ around the axis
${\bf\Delta}_{RX}=-
\vec q_{RX}$ we have
\be\label{D.2}
{\cal M}^f_{ns}={\cal M}^{f\chi}_{ns} \, e^{i\chi(s-n)},\ee
where $f\chi$ denotes the rotated final state $f$. This leads to
\bear\label{D.3}
S_{mr,ns}&=&\frac{1}{2m_p}\sum_{f\in X}
\delta(\Delta+q-p_f){\cal M}^{f\ *}_{mr} {\cal M}^f_{ns}\nonumber\\
&=&\frac{1}{2m_p}\sum_{f\in X}\delta(\Delta+q-p_f)
{\cal M}^{f\chi *}_{mr}{\cal M}^{f\chi}_{ns}\nonumber\\
&&\cdot \exp [i\chi(s-n)-i\chi(r-m)].\ear
Now we use in an essential way that with our criteria  I or II
the set of states $X$ over which we sum in (\ref{D.3}) is invariant
under the rotation $R(\chi)$. Then the summation over $f$ is identical
to the summation over $f\chi$ and we get from (\ref{D.3}):
\be\label{D.4}
S_{mr,ns}=S_{mr,ns}\exp[i\chi(s-n-r+m)].\ee
Since $\chi$ is arbitrary, this implies (\ref{5.12}), q.e.d.

\newpage

\section*{Table Captions}
\begin{description}
\item{Table 1} Definitions of kinematic variables for $ep$
scattering (\ref{1.1}).

\item{Table 2} Useful relations for the kinematic variables of
$ep$ scattering (\ref{1.1}).

\item{Table 3} The matrix ${\cal M}$ (\ref{4.11}) relating the cross sections 
and interference terms $\tilde\sigma_{mn}^{(X)}$ (\ref{4.10})
to $\sigma_{mn}^{(X)}$ (\ref{3.13}) where the relations (\ref{3.26})
are assumed to hold.
\end{description}

\section*{Figure Captions}
\begin{description}
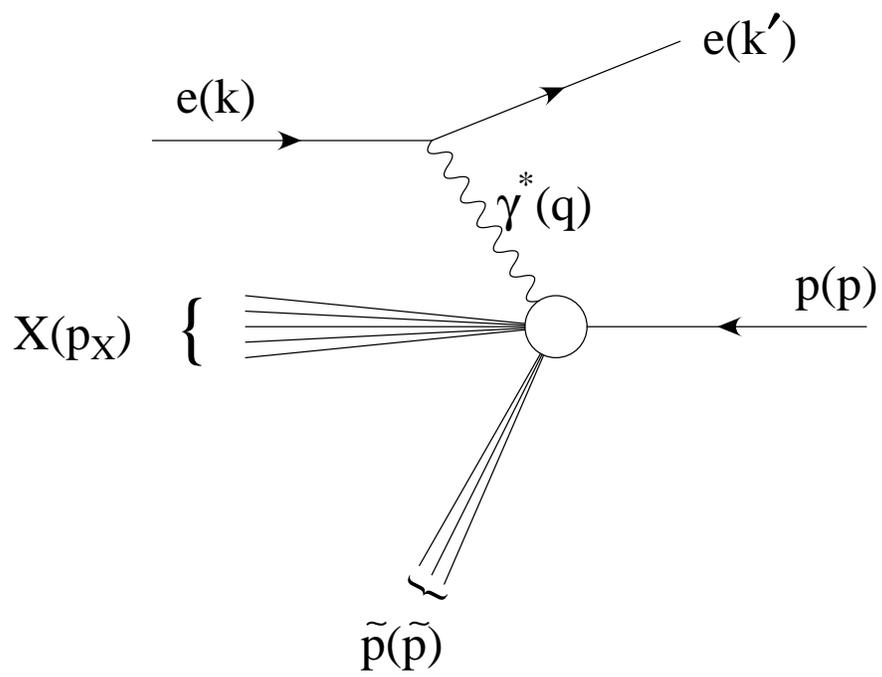
\item{Fig.~1} Diagram for $ep$ scattering (\ref{1.1})
in the one-photon exchange approximation $(q=k-k')$.

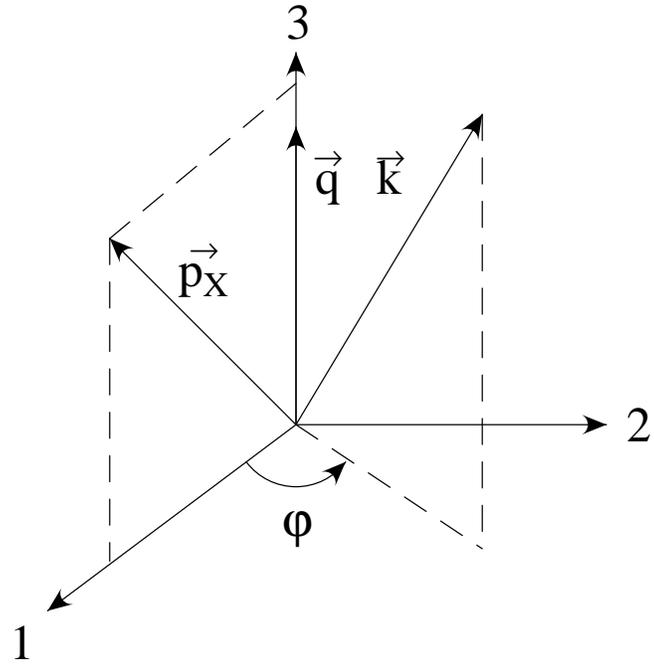
\item{Fig.~2} The coordinate system chosen in the
rest system of the original proton and the definition
of the angle $\varphi$.

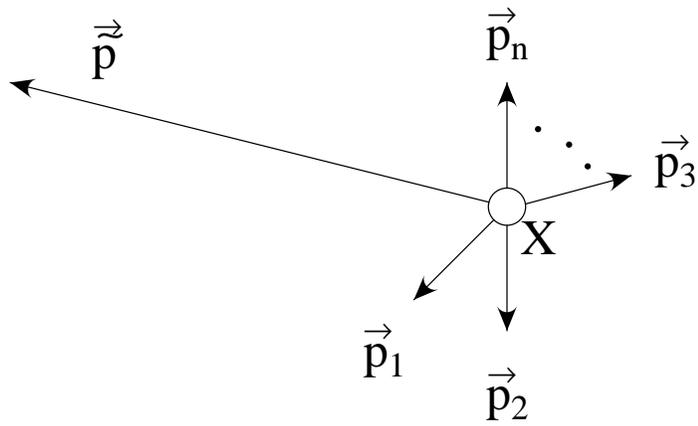
\item{Fig.~3} A typical final state allowed by criterion
II as rapidity gap event. The momentum configuration is
drawn in the $X$ rest system where we require $p_i^0\leq
M^2_{cut}/m_X$ for all $i$.

\item{Fig.~4} The definition of the azimuthal angles
$\tilde\psi,\psi'$ and $\varphi'$ in the 1-2 plane of
the HERA coordinate system, where the 3-axis is in the
direction of flight of the original proton.

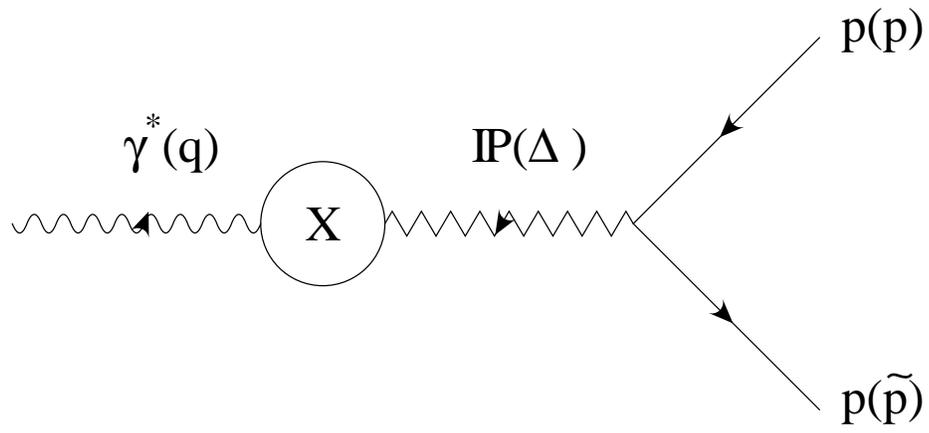
\item{Fig.~5} The reaction $\gamma^*p\to pX$ in the rest
system of $X(RX)$. In models where this reaction proceeds via
Pomeron exchange the virtual photon and the Pomeron collide
head on in this system.

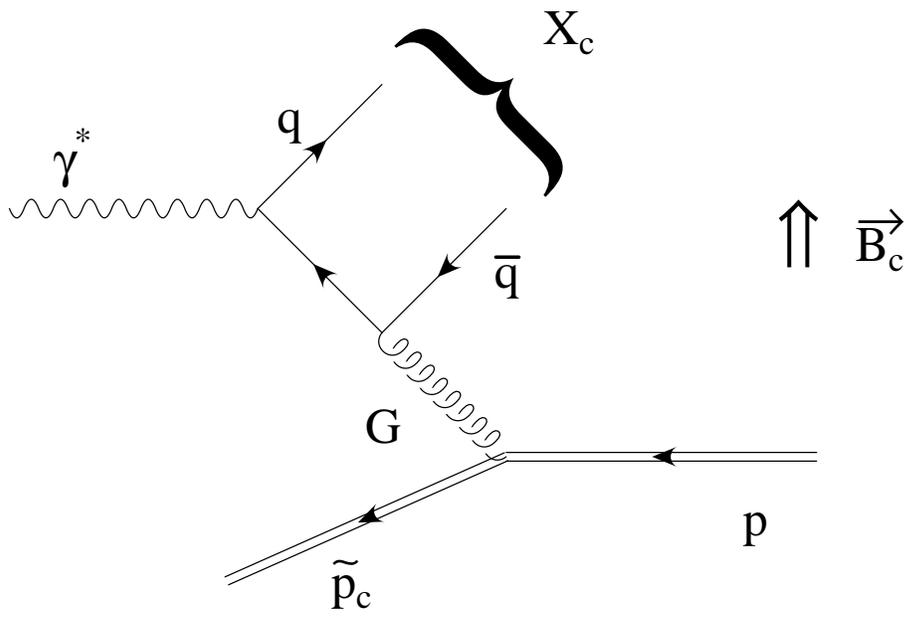
\item{Fig.~6} The reactions (\ref{5.59}), (\ref{5.60})
taking place in a QCD vacuum background colour field, here
a chromomagnetic field $\vec B_c$.

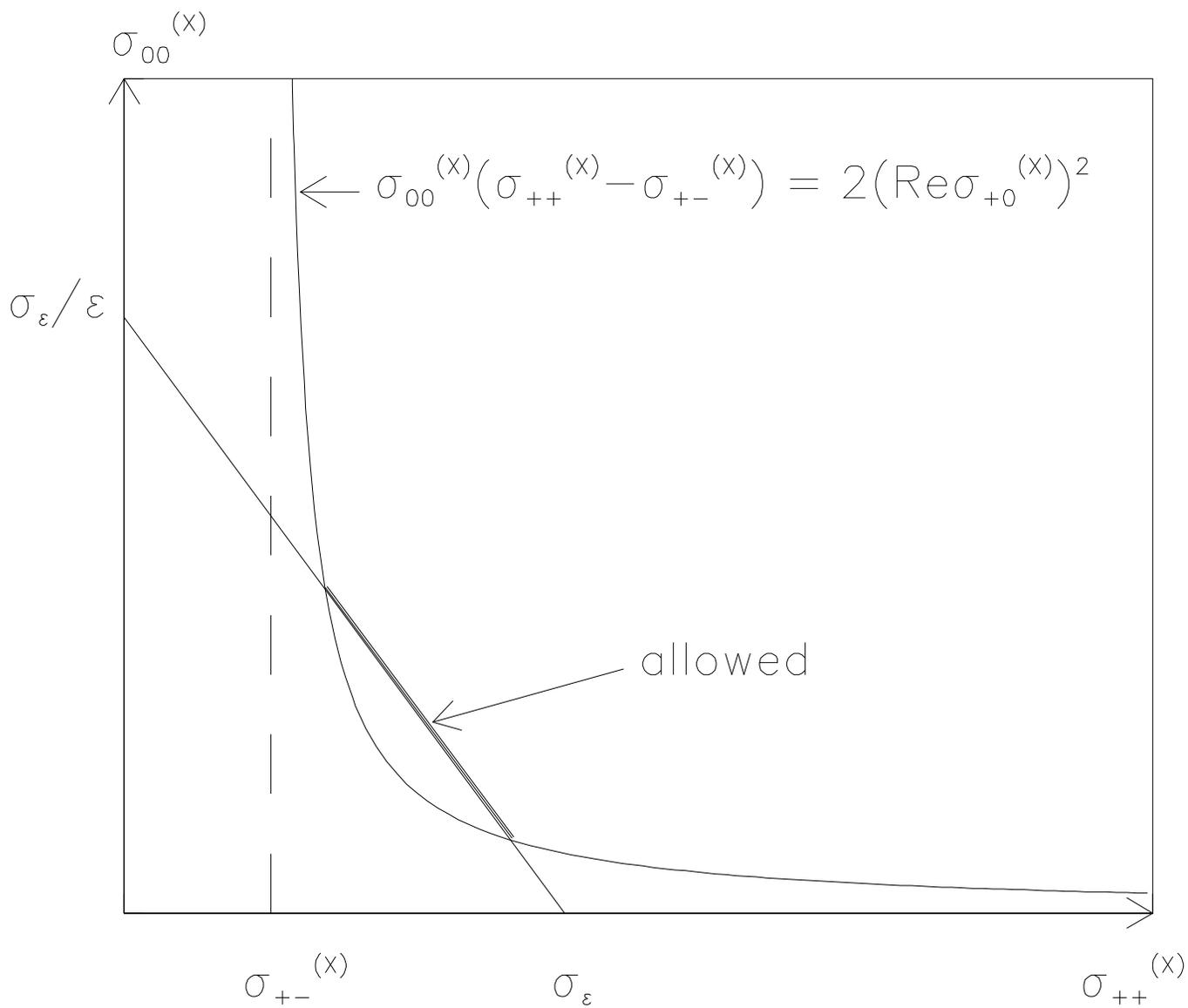
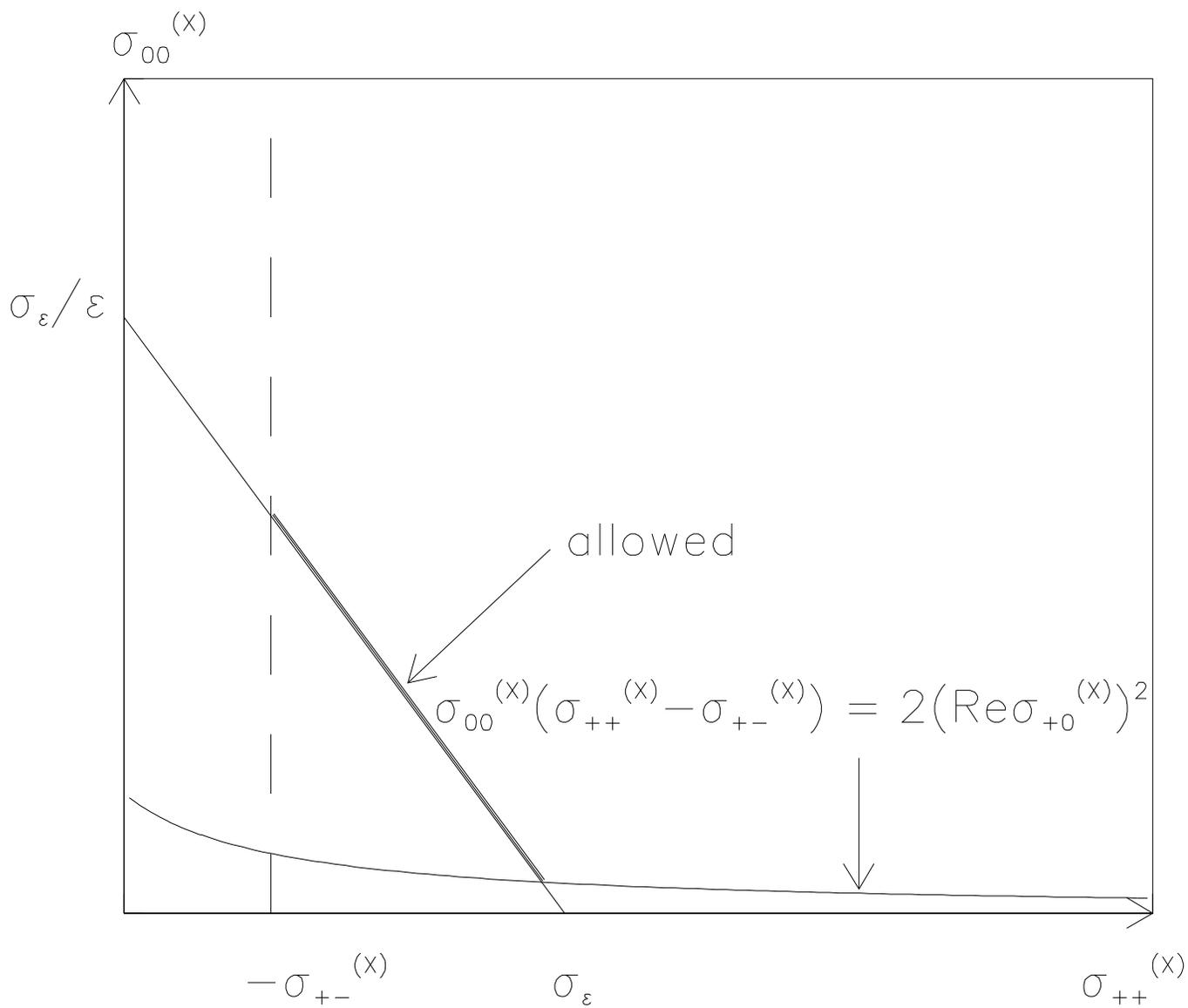
\item{Fig.~7} The allowed range (thick line) for $\sigma_{++}^{(X)}$ and
$\sigma_{00}^{(X)}$ if $\sigma_{\varepsilon}$ (\ref{C.7}) and the
interference terms $\sigma_{+-}^{(X)}$ and ${\rm Re}
\sigma_{+0}^{(X)}$ have been
measured. The case $\sigma_{+-}^{(X)}\ge 0$ corresponds to (a), the case
$\sigma_{+-}^{(X)}<0$ to (b).

\end{description}

\newpage
\begin{center}
{\bf Table 1}
\end{center}
\begin{eqnarray*}
&&q=k-k',\quad \tilde m^2=\tilde p^2,\\
&&\Delta=p-\tilde p,\quad m^2_X=p_X^2;\\
&&\\
&&s=(p+k)^2;\\
&&W^2=(p+q)^2,\quad K=\frac{W^2-m_p^2}{2m_p},\\
&&t=(p-\tilde p)^2,\\
&&u=(p-p_X)^2;\\
&&Q^2=-q^2,\quad\nu=\frac{p\cdot q}{m_p},\\
&&x=\frac{Q^2}{2p\cdot q},\quad y=\frac{p\cdot q}{p\cdot k}\\
&&\varepsilon=\frac{2(1-y)-2 xy\, m^2_p(s-m^2_p)^{-1}}{1+(1-y)^2+2 xy\,
 m_p^2(s-m^2_p)^{-1}};\\
&&\xi=\frac{\Delta \cdot q}{p\cdot q},\\
&&\beta=\frac{Q^2}{2\Delta \cdot q}.\end{eqnarray*}

\newpage
\begin{center}
{\bf Table 2}
\end{center}
\begin{eqnarray*}
p+q&=&\tilde p+p_X,\\
\Delta+q&=&p_X,\\
W^2+t+u&=&m_p^2+\tilde m^2+m_X^2-Q^2,\\
2p\cdot k&=&s-m^2_p,\\
2p\cdot q&=&y(s-m_p^2),\\
Q^2&=&xy(s-m^2_p),\\
2p_X\cdot q&=&m_X^2-t-Q^2=-xy(s-m^2_p)+m_X^2-t,\\
2p_X\cdot p&=&y(s-m^2_p)+t+m_p^2-\tilde m^2,\\
(p\cdot q)^2+Q^2m^2_p&=&m_p^2(\nu^2+Q^2)\\
&=&m^2_p\nu^2\left[1+\frac{4x^2m^2_p}{Q^2}\right]\\
&=&\frac{1}{4}y(s-m^2_p)^2\left[y+\frac{4xm^2_p}{s-m^2_p}\right]\\
\xi&=&\frac{m^2_X+Q^2-t}{W^2+Q^2-m^2_p},\\
\beta&=&\frac{x}{\xi},\quad 0\leq\beta\leq1;\\
\beta&=&\frac{Q^2}{m^2_X+Q^2-t},\\
m_X^2&=&\frac{Q^2(1-\beta)+t\beta}{\beta},\\
\Delta_{RX}^0&=&\frac{\Delta \cdot p_X}{m_X}=\frac{m_X^2+Q^2+t}{2m_X}\\
&=&\frac{Q^2}{2\beta m_X}\left(1+\frac{2\beta t}{Q^2}\right)\\
&=&\frac{1}{2}Q\left[\beta(1-\beta)+\frac{\beta^2t}{Q^2}\right]^{-1/2}\left(1+
\frac{2\beta t}{Q^2}\right),\\ 
|{\bf\Delta}|_{RX}&=&\frac{1}{2} Q\left[\beta(1-\beta)+\frac{\beta^2t}
{Q^2}\right]^{-1/2}\left(1+\frac{4\beta^2t}{Q^2}\right)^{1/2}\end{eqnarray*}

\newpage
\begin{center}
{\bf Table 3}
\end{center}
\begin{tabular}{l|ccccc}
&$\sigma_{++}^{(X)}$&$\sigma_{00}^{(X)}$&
$\sigma_{+-}^{(X)}$&${\rm Re}\sigma_{+0}^{(X)}$&
${\rm Im}\sigma_{+0}^{(X)}$\\
\hline\\
$\tilde\sigma_{++}^{(X)}$&$(\gamma_X^2+1)/2$&
$v_X^2\gamma_X^2/2$&$-v_X^2\gamma_X^2/2$&
$-\sqrt2 v_X\gamma_X^2$& 0 \\
$\tilde\sigma_{00}^{(X)}$&$v_X^2\gamma_X^2$&
$\gamma^2_X$&$-v_X^2\gamma_X^2$&$-2\sqrt2 v_X\gamma_X^2$&0\\
$\tilde\sigma_{+-}^{(X)}$&$-v_X^2\gamma_X^2/2$&
$-v_X^2\gamma_X^2/2$&$(\gamma_X^2+1)/2$&
$\sqrt2 v_X\gamma_X^2$& 0 \\
${\rm Re}\tilde\sigma_{+0}^{(X)}$&$v_X\gamma_X^2/\sqrt2$&
$v_X\gamma_X^2/\sqrt2$&$-v_X\gamma_X^2/\sqrt2$&
$-(1+v_X^2)\gamma_X^2$& 0 \\
${\rm Im}\tilde\sigma_{+0}^{(X)}$&0&0&0&0&$-1$ \\
\end{tabular}

\end{document}